%% file: MultipairFullDuplexRateSplittingTWC_v2.tex
\documentclass[twocolumn,10pt]{IEEEtran}

\input{def.tex}

\DeclareSymbolFont{matha}{OML}{txmi}{m}{it}
\DeclareMathSymbol{\varv}{\mathord}{matha}{118}
\usepackage{amsmath}
\usepackage{eurosym}

\newcommand{\overbar}[1]{\mkern 1.5mu\overline{\mkern-1.5mu#1\mkern-1.5mu}\mkern 1.5mu}

\begin{document}
\title{Rate-Splitting Robustness in Multi-Pair Massive MIMO Relay Systems}\author{Anastasios Papazafeiropoulos  and Tharmalingam Ratnarajah    \\
\thanks{A. Papazafeiropoulos was with the  Institute for Digital Communications (IDCOM), University of Edinburgh, Edinburgh, EH9 3JL, U.K. He is now with the University of Hertfordshire, Hatfield, AL10 9AB,   U.K.  T. Ratnarajah is  with the  Institute for Digital Communications (IDCOM), University of Edinburgh, Edinburgh, EH9 3JL, U.K., (email: tapapazaf@gmail.com, t.ratnarajah@ed.ac.uk). }
\thanks{This work was supported by the U.K. Engineering and Physical Sciences Research Council (EPSRC) under grant EP/N014073/1.}
\thanks{Parts of this work were presented in IEEE Global Commun.Conf. (GLOBECOM) 2017, Singapore, Dec. 2017~\cite{Papazafeiropoulos2017Globecom}.}}
\maketitle\vspace{-4mm}
\begin{abstract}
Relay systems improve both coverage and system capacity. Towards this direction, full-duplex (FD)  technology, being able to boost the spectral efficiency by transmitting and receiving simultaneously on the same frequency and time resources, is envisaged to play a key role in future networks. However, its benefits come at the expense of self-interference (SI) from their own transmit signal. At the same time, massive multiple-input massive multiple-output (MIMO) systems, bringing unconventionally many antennas, emerge as a promising technology with huge degrees-of-freedom (DoF). To this end, this paper considers a multi-pair decode-and-forward FD relay channel, where the relay station is deployed with a large number of antennas. Moreover, the rate-splitting (RS) transmission has recently been shown to provide significant performance benefits in various multi-user  scenarios with imperfect channel state information at the transmitter (CSIT). Engaging the RS approach, we employ the deterministic equivalent (DE) analysis to derive the corresponding sum-rates in the presence of interferences.  Initially, numerical results demonstrate the robustness of RS in half-duplex (HD) systems, since the achievable sum-rate increases without bound, i.e., it does not saturate at high signal-to-noise ratio (SNR). Next,  we tackle the detrimental effect of SI in FD. In particular, and most importantly, not only FD outperforms HD, but also RS enables  increasing the range of SI over which FD outperforms HD.  Furthermore, increasing the number of relay station antennas, RS appears to be more  efficacious due to imperfect CSIT, since SI decreases. Interestingly, increasing the number of users, the efficiency of RS worsens and its implementation becomes less favorable under these conditions. Finally, we verify that the proposed DEs, being accurate for a large number of relay station antennas, are tight approximations even for realistic system dimensions.
\end{abstract}

\section{Introduction}
Massive multiple-input massive multiple-output (MIMO) technology is a key enabler for the fifth generation (5G) wireless communication systems  achieving energy-efficient transmission and high spectral efficiency~\cite{METIS,Rusek2013,Larsson2014}. According to its characteristic topology, a large number of service antennas per unit area performs coherent linear processing, and offers an unprecedented number of degrees-of-freedom (DoF). Among its benefits, we emphasize the substantial  reduction of both intra-cell and inter-cell interference, which ultimately, lead to high performance efficiency, both spectral and energy.

In a parallel direction, in-band full-duplex (FD) is a novel technology that doubles the throughput induced by standard half-duplex relaying by means of simultaneous transmission and reception at the same frequency and time during a wireless communication~\cite{Bliss2007,Bliss2012}.  Moreover, its theoretical and experimental progress towards its practical implementation~\cite{Riihonen2011a,Zheng2013,Duarte2012} is notable. Actually, the theoretical progress can lead to a practical achievement with new opportunities. However,  this is quite demanding because the FD transmission is accompanied by an inherent obstacle. Specifically, this obstacle is the so-called self-interference (SI) due to the leakage from the relay's output to its input~\cite{Riihonen2011a}. It is worthwhile to mention that the main difference between SI and general interference is that SI is known at the receiver, which could be sufficient  for SI suppression. There are
several challenges for the mitigation of SI, being crucial for FD operation. For example, the received signal and the SI may exhibit a large amplitude difference going to exceed the dynamic range of the analog-to-digital converter at the receiver side~\cite{Duarte2012}.   Although the SI cancellers try their best to maximize the cancellation performance, residual interference remains and rate saturation at high signal-to-noise ratio (SNR) appears.  Hence, the circumvention of the harmful consequences of the SI takes a prominent position in the research area of FD systems. Among the suppression methods for SI, MIMO processing, specialized in the spatial domain, provides an exceptionally effective means~\cite{Riihonen2011a,Riihonen2011,Sung2011}. As a result, driving to massive MIMO is a reasonable approach for next-generation systems.


To grasp the  benefits of massive MIMO   the accurate knowledge  of channel state information at the transmitter (CSIT) is required. In fact, accurate CSIT becomes even more challenging  as the number of antennas increases~\cite{Andrews2014,Lu2014}. In such case, the  Time Division Duplex (TDD) design has proved to be a more feasible solution against Frequency Division Duplex (FDD) schemes because the latter are accompanied with further channel estimation and feedback challenges~\cite{Marzetta2010,Rusek2013,Papazafeiropoulos2015a,Larsson2014,Hoydis2013}. The DoF decrease as the CSIT inaccuracy increases. Especially, in realistic scenarios, where CSIT is imperfect, linear precoding techniques lead to a rate ceiling at high SNR, if the error variance is fixed. 

In order to enhance the sum DoF, the rate splitting (RS) strategy has been proposed~\cite{Yang2013}.  The RS outperforms conventional broadcasting  at high SNR because it does not experience any ceiling effect~\cite{Hao2015,Dai2016,Papazafeiropoulos2017,Papazafeiropoulos2017a}\footnote{Interestingly, a further gain of RS over no RS (NoRS)  can be achieved by optimizing the precoders~\cite{Joudeh2016}, where for the sake of exposition and comparison, henceforth, we denote by NoRS all the  conventional techniques to contrast with the RS techniques.}. According to this strategy, the message intended for one user is split into a private part and a common part by using a fraction of the total power. The private part is transmitted by means of zero-forcing (ZF) beamforming, while the common part is superimposed on top of the precoded private part by means of the remaining power. The common message is drawn from a public codebook and decoded by all users. At the receiver side, the decoding procedure involves first the decoding of the common message by means of successive interference cancellation, and then, the decoding of the private message of each user follows.  

Although the relaying in previous cellular generations was mostly used for coverage enhancement,  in today's cellular networks, it is shown that it can improve both coverage and system capacity~\cite{Chung2007}. In this regard,   relaying has been already considered as one of the salient features in 3GPP Long Term Evolution (LTE) advanced \cite{Hoymann2012}.  Especially, the importance of relaying in massive MIMO systems has been already demonstrated in several studies~\cite{Chen2016} .

In the area of massive MIMO relaying,  both half-duplex (HD) and FD have been studied~\cite{Cui2014,Xiong2016,Sun2016,Xia2015,Chen2016a,Ngo2014,Xia2015a}. In particular, in the case of HD, the spectral efficiency has been investigated for a very large number of relay station antennas \cite{Suraweera2013,Cui2014,Xiong2016,Sun2016,Xia2015,Chen2016a}. On the other hand, e.g., FD relaying with a large number of antennas and linear processing as well as the scaling behavior with the number of relay antennas of the self-interference were analyzed in~\cite{Ngo2014} in terms of the end-to-end achievable rate. Towards this direction, the asymptotic performance of amplify-and-forward massive MIMO relay systems with additive hardware impairments was determined in~\cite{Xia2015a}.
\subsection{Motivation-Contributions}
Following the research trends and needs in massive MIMO and FD systems, we consider  a collection of $K$ sources communicating with another collection of $K$ destinations through  an intermediate  massive MIMO FD relay station, and we focus on the application of RS. In particular, in our architecture scenario, two sources, leading to rate saturation, are faced. The first includes the multi-user interference with imperfect CSIT in the second link, and the second concerns the  SI emerging from the FD transmission. This work tackles the challenge of mitigating the rate saturation by leveraging the RS approach. In particular, we investigate the robustness of the RS method in realistic massive MIMO FD settings suffering from both pilot contamination and SI. The motivation of this work started by the observation that in FD systems the CSIT is altered due to the presence of SI. Furthermore, it is known that RS is applicable in multi-user settings with imperfect CSIT. Hence, these observations suggest that RS will be effective in the mitigation of the SI and the consecutive circumvention of the rate saturation due to the overall imperfect CSIT. Note that our system setup is quite general, since it can model cellular networks with  some users transmitting  simultaneously
signals to several other users via an infrastructure-based relay station serving  several roles such as a low power base station~\cite{Yang2009}. Moreover, having a MIMO relaying in the scene, we test RS in the basic scenario of just HD transmission. The main contributions are summarized as follows:
\begin{itemize}
\item Contrary to existing works such as~\cite{Bliss2007,Bliss2012,Riihonen2011a,Zheng2013,Duarte2012,Riihonen2011,Sung2011}, which have studied  FD MIMO systems, we focus on massive MIMO systems, and examine the impact of SI, when RS transmission is applied at the second link. For the sake of comparison, we also present the results corresponding to an HD relay system. It is shown that RS is robust in both multipair HD and FD settings.
\item We derive the deterministic SINRs of NoRS and RS in multipair FD systems with imperfect CSIT and use them to investigate the performance benefits of RS over NoRS in the presence of SI. Actually, first, we obtain the estimated channels of both links by means of MMSE estimation. Next, we apply RS in the second link by designing the precoder of the  private and common messages, and we consider suitable power allocation. Although the basic implementation of the RS strategy assumes just ZF precoding for the transmission of the private messages except~\cite{Yang2013}, we consider regularized ZF (RZF) precoding because it is another low-complexity  linear processing technique applicable in massive MIMO systems. However, RZF provides  better performance than ZF. Finally, we provide the DEs of the SINRs  of the  private and common messages. Note that these deterministic expressions allow avoiding any Monte Carlo simulations with very high precision. 
\item   Above this, RS is robust in HD and FD scenarios because it can mitigate the multi-user interference taking place in the second link of both HD and FD cases. In fact, we elaborate on the impact of the severity of SI. Actually, RS is able to mitigate the saturation due to  the SI in spite of the knowledge of perfect or imperfect CSIT. Furthermore, in the case of lower SI, RS behaves better. The same observation is made as the number of relay station antennas is increased, since then, SI becomes lower.
 \item We show that an increase of the  number of user elements (UEs) in a multipair FD system results in a reduction of the performance gain of RS over NoRS because the common message has to be decoded by more UEs. Moreover, we quantify this decrease exhibited due to a less mitigated SI.
\end{itemize}

The   remainder of this paper is structured as follows.  Section~\ref{SystemModel} presents the system  and signal models for both links of the multi-pair  FD relay system.   Section~\ref{DataTransmission} presents the data transmission phase, while in Section~\ref{ChannelEstimation}, we provide the estimated channels obtained during the uplink training phase of the two links. Next, we present the RS approach. In Section~\ref{EndtoEndTransmission}, we  present the end-to-end transmission by obtaining the SINR of each link. Section~\ref{Achievable} exposes the DE analysis, which enables the design of the  precoder of the common message, and mainly, the derivation of the achievable rates in the presence of SI.   The numerical results are placed in Section~\ref{NumericalResults}, while Section~\ref{Conclusions} summarizes the paper.

\begin{table*}
\caption{Notations Summary}
\begin{center}
                                             \begin{tabulary}{\columnwidth}{ | c | c | }\hline
{\bf Notation} &{\bf Description}\\ \hline
$K$ & Communication pairs \\ \hline
$M$, $N$ & Numbers of transmit and receive antennas\\ \hline
$\mathrm{S}_{k}$, $\mathrm{D}_{k}$ & The $k$th source and destination\\ \hline
$T_{c}$,  $B_{c}$ & Coherence time, bandwidth\\ \hline
$\sigma^{2}_{\mathrm{SI}}$  & The variance of the elements of the self-interference matrix\\ \hline
$\tau$& Duration of the training phase
\\ \hline
$p_{\mathrm{S}}$, $p_{\mathrm{tr}}$ & Average transmit power per source and transmit power per pilot symbol\\ \hline
$\bG_{\mathrm{SR}}$, $\bG_{\mathrm{rD}}$, $\bG_{\mathrm{RR}}$  & Channel matrices of the first link, second link, and self-interference\\ \hline
$\bH_{\mathrm{SR}}$, $\bH_{\mathrm{RD}}$  & Small-scale fading matrices of the first and second links \\ \hline
$\bD_{\mathrm{SR}}$, $\bD_{\mathrm{RD}}$  & Large-scale fading matrices of the first and second links\\ \hline
$\bff_{c}$, $\bff_{k}$ & Precoding vectors of the common and private messages corresponding to UE $k$ \\ \hline
$\rho_{c}$, $\rho_{k}$ & Powers allocated to the common and private messages corresponding to UE $k$\\ \hline
$\lambda$ & Normalization of the precoded message\\ \hline
$\gamma_{\mathrm{SR}}$, $R_{\mathrm{SR}}$  & SINR and achievable rate of the first link \\ \hline
$ \gamma_{\mathrm{RD},k}^{\mathrm{c}}$, $ \gamma_{\mathrm{RD},k}^{\mathrm{p}}$  & SINR of the common and private messages of the second link \\ \hline
$ R_{\mathrm{RD}}^{\mathrm{c}}$, $ R_{\mathrm{RD},k}^{\mathrm{p}}$  & Achievable rates of the common and private messages of the second link \\ \hline
\end{tabulary}
                                            \end{center}
\end{table*}

\textit{Notation:} Vectors and matrices are denoted by boldface lower and upper case symbols. $(\cdot)^\T$, $(\cdot)^*$,  $(\cdot)^\H$, and $\tr\!\left( {\cdot} \right)$ represent the transpose, conjugate, Hermitian  transpose, and trace operators, respectively. The expectation  operator is denoted by $\EE\left[\cdot\right]$. The $\mathrm{diag}\{\cdot\}$ operator generates a diagonal matrix from a given vector, and the symbol $\triangleq$ declares definition. The notations $\mathbb{C}^{M \times 1}$ and $\mathbb{C}^{M\times N}$ refer to complex $M$-dimensional vectors and  $M\times N$ matrices, respectively. Finally, $\bb \sim \cC\cN{(\b0,\mathbf{\Sigma})}$   denotes a circularly symmetric complex Gaussian variable with zero-mean and covariance matrix $\mathbf{\Sigma}$.
\section{System Model}\label{SystemModel}
The concept of our model involves a multipair FD relaying system with a common relay station $\mathrm{R}$ and $K$ communication pairs $\left( \mathrm{S}_{k},\mathrm{D}_{k}\right),~k=1,\ldots,K$ sharing the same time-frequency resources. Specifically, we consider $K$ user pairs, where the $k$th source $\mathrm{S}_{k}$ exchanges information through a relay operating in decode-and-forward protocol with the $k$th UE destination $\mathrm{D}_{k}$. Moreover, the system suffers from SI  due to the simultaneous transmission and reception, since it operates under an FD mode. Note that there is no direct link between the source $\mathrm{S}_{k}$ and the corresponding destination $\mathrm{D}_{k}$ because of heavy shadowing and large path-loss. The source and the destination pairs are equipped with a single antenna, while the FD relay station is deployed with $N$  receive antennas and $M$ transmit antennas, i.e., it includes $V=M+N$ antennas in total\footnote{This network configuration is of high  practical interest. For example, it can describe a cellular setup, where the  communication between two users is performed by means of a massive antenna low power base station.}. 
\subsection{Signal Model}
We consider frequency-flat channels between the source user $k$ and the relay as well as between the relay and destination UE $k$, modeled as Rayleigh block fading.  The channels are  assumed static across a coherence block of $T$ channel uses with the channel realizations between blocks being independent. The size of the block is defined by the product between the coherence time $T_{c}$ and the coherence bandwidth $B_{c}$.  Specifically, the frequency-flat channel matrices between the $K$ sources and the relay station's receive antenna array as well as between the relay station's transmit antenna array and the $K$ destinations, modeled as Rayleigh block fading, are denoted by $\bG_{\mathrm{SR}}\in \bbC^{N \times K}$ as well as $\bG_{\mathrm{RD}}\in \bbC^{M\times K}$, respectively. We express each channel realization as\footnote{According to  the favorable propagation assumption which has been validated in practice~\cite{Duarte2012}, we consider that the channels from the relay station to different sources and destinations are independent~\cite{Ngo2014}.}
\begin{align}
\bG_{\mathrm{SR}}&\triangleq \bH_{\mathrm{SR}} \bD_{\mathrm{SR}}^{1/2}\\
\bG_{\mathrm{RD}}&\triangleq \bH_{\mathrm{RD}} \bD_{\mathrm{RD}}^{1/2}.
\end{align}

These channel matrices account for both small-scale  and large-scale fadings. Specifically, the  matrices $\bH_{\mathrm{SR}}$ and $\bH_{\mathrm{RD}}$, having independent and identically
distributed (i.i.d.) $\mathcal{CN}\left( 0,1 \right)$ elements, describe small-scale
fading, while the matrices $\bD_{\mathrm{SR}}$ and $\bD_{\mathrm{RD}}$ are diagonal and express the large-scale fading  in terms of the $k$th
diagonal elements, which are denoted by $\beta_{\mathrm{SR},k}$ and $\beta_{\mathrm{RD},k}$, respectively. Furthermore, assuming that there is no line-of-sight component, the SI channel is modeled by means of  the Rayleigh fading distribution. Mathematically, it is described by the $\bG_{\mathrm{RR}}\in \bbC^{M\times N}$ channel matrix between the relay's transmit and receive arrays. In other words, the elements of the SI channel matrix $\bG_{\mathrm{RR}}$ can be modeled as i.i.d. complex Gaussian random variables with zero mean and variance $\sigma^{2}_{\mathrm{SI}}$, i.e., $\mathcal{CN}\left( 0,\sigma^{2}_{\mathrm{SI}} \right)$. The physical meaning of $\sigma^{2}_{\mathrm{SI}}$ can be seen as 
the level of SI that is dependent on the distance
between the transmit and receive antenna arrays. Also, the assumption that the channels between the transmit and receive antennas are i.i.d.  considers that the distance between the transmit and the receive arrays is much larger than the distance between the antenna elements. 
\section{End-to-End Transmission}\label{DataTransmission}
This section presents the data transmission and the uplink estimation phases of the multipair decode-and-forward FD model as well as the RS approach. 
\subsection{Data Transmission}
At time instant $n$, the  $K$ user sources $\mathrm{S}_{k}$ ($k=1,\ldots,K$) transmit simultaneously their signals to the relay, which, in turn, broadcasts the signal to all $K$ destinations. Actually, we denote $\sqrt{p_{\mathrm{S}}}u_{k}\left[n \right] $ the $k$th user transmit signal at  time $n$ with $p_{\mathrm{S}}$  being the average transmit power
of each source since $\EE \left\{|u_{k}\left[ n \right]|^{2} \right\}=1 $, while at the relay station the received signal is interfered with its transmit signal.

Herein, we present the conventional input-output signal model (NoRS) as a measure of comparison. More precisely, the  signal received by the receive antenna array of the relay    from all the sources is given by~\cite{Ngo2014}
\begin{align}
\by_{\mathrm{R}}\left[ n\right] =\sqrt{p_{\mathrm{S}}}\bG_{\mathrm{SR}}\bu\left[n \right] +\bG_{\mathrm{RR}}\bs\left[n \right]+ \bz_{\mathrm{R}}\left[ n\right], \label{BasicSystemModelwithoutRS1}
\end{align}
while the signal received by the $K$ destinations from the transmit antenna array of the relay station   is written as
\begin{align}
\by_{\mathrm{D}}\left[ n\right] =\bG_{\mathrm{RD}}^{\H}\bs\left[n \right]+ \bz_{\mathrm{D}}\left[ n\right], \label{BasicSystemModelwithoutRS2}
\end{align}
where $\bz_{\mathrm{R}} \sim \cC\cN(0,\Id_{N})$ and $\bz_{\mathrm{R}} \sim \cC\cN(0,\Id_{K})$ are the additive white Gaussian noises (AWGNs) at the relay station and
the $K$ destinations, respectively. Note that  $u[n]$ is a vector whose k-th element is $u_k[n]$, and the vector $s[n]$ expresses the transmitted signal from relay to destinations. For the sake of complexity, we assume that the relay station applies linear processing, i.e., the relay station achieves the decoding of the transmitted signals from the $K$ sources by employing a linear receiver, and at the same time, the relay forwards the signals to the $K$ destinations by using  linear
precoding. In the general case, the linear decoder and precoder are given by $\bW^{\H}$ and $\bF$, respectively. Specifically, the received signal is seperated into $K$ streams after multiplication with the linear receiver $\bW^{\H}$ according to
\begin{align}
\br\left[ n\right]&=\bW^{\H}\by_{\mathrm{R}}\left[ n\right]\nn\\
&=\sqrt{p_{\mathrm{S}}}\bW^{\H}\bG_{\mathrm{SR}}\bu\left[n \right] +\bW^{\H}\bG_{\mathrm{RR}}\bs\left[n \right]+\bW^{\H} \bz_{\mathrm{R}}\left[ n\right]. \label{BasicSystemModelwithoutRS3}
\end{align}

The $k$th element of $\br\left[n\right]$, or equivalently, the $k$th stream enables the decoding of the signal transmitted from the $k$th source  $\mathrm{S}_{k}$. More precisely, we have
\begin{align}
\br_{k}\left[ n\right]&=\sqrt{p_{\mathrm{S}}}\bw_{k}^{\H}\bg_{\mathrm{SR},k}u_{k}\left[n \right] +\sqrt{p_{\mathrm{S}}}\sum_{j\ne k}^{K}\bw_{k}^{\H}\bg_{\mathrm{SR},j}u_{j}\left[n \right]\nn\\
&+\bw_{k}^{\H}\bG_{\mathrm{RR}}\bs\left[n \right]+\bw_{k}^{\H} \bz_{\mathrm{R}}\left[ n\right], \label{BasicSystemModelwithoutRS4}
\end{align}
where the first and second terms represent the desired signal and the interpair interference, while the third and last term express the SI and the post-processed noise. Note that $\bg_{\mathrm{SR},k}$ and $\bw_{k}$ are the $k$th columns of $\bG_{\mathrm{SR}}$ and $\bW$, respectively.

Having detected the signals transmitted from the $K$ sources, the relay station employs linear precoding to process them. Then, the relay station broadcasts the signals to all $K$ destinations. If we assume that the processing delay is equal to $d\ge 1$,  we have\footnote{This common assumption in the  existing literature for FD systems, enables us to assume that at a given time instant, the receive and transmit signals at the relay station are uncorrelated. Also, we assume that the relay can obtain the source signals without any error. Otherwise, $s[n]$ in~\eqref{precodesSignal} would include a noise term.}
\begin{align}
\bs\left[ n\right]= \bu\left[n-d \right],\label{precodesSignal} 
\end{align}
where $ \bu\left[n-d \right]$ includes the linear precoding matrix. By substituting of~\eqref{precodesSignal} into~\eqref{BasicSystemModelwithoutRS2}, we obtain the received signal at $\mathrm{D}_{k}$ as
\begin{align}
 y_{\mathrm{D},k}\left[ n\right] &=\bg_{\mathrm{RD},k}^{\H}\bu\left[n-d \right]+z_{\mathrm{D},k}\left[ n\right]
 \label{BasicSystemModelwithoutRS5}
\end{align}
with $\bg_{\mathrm{RD},k}$  being the $k$th column of $\bG_{\mathrm{RD}}$, while $z_{\mathrm{D},k}$ is the $k$th element of $\bz_{\mathrm{D}}$.

Choosing MMSE/RZF processing, i.e., employing MMSE for the decoder and RZF for the precoder, we achieve to maximize the received SNR by not taking into account the interpair interference\cite{Hoydis2013}. In other words, MMSE and RZF behave quite well.
Hereafter, we omit the time index from our analysis for the sake of simplicity.
 \subsection{Pilot Training Phase}\label{ChannelEstimation}
In practical systems, the relay station has to estimate both the channels $\bG_{\mathrm{SR}}$ and $\bG_{\mathrm{RD}}$. A good transmission protocol to implement the current design is TDD, which is the most favorable scheme for massive MIMO. According to TDD, the protocol consists of coherence blocks having duration of $T$ channel uses. In turn, each block is split into $\tau \ge 2K$ training pilot symbols to guarantee that the source and the destination user elements (UEs) are spatially separable by the relay station and the remaining channel uses are allocated for the  data transmission symbols\footnote{The pilot sequences of $\tau$ symbols are transmitted simultaneously  by all the sources and destinations.}. Note that during the data transmission phase, the channel is known due to the property of the channel reciprocity. After sending the pilots, the received signal matrices at the receive and transmit antennas of the relay are given by
\begin{align}
\bY_{\mathrm{r}}^{\mathrm{tr}}&=\sqrt{\tau p_{\mathrm{tr}}}\left( \bG_{\mathrm{SR}}\bPhi_{\mathrm{S}}+\bar{\bG}_{\mathrm{RD}}\bPhi_{\mathrm{D}} \right)+\bZ^{\mathrm{tr}}_{\mathrm{r}},\\
\bY_{\mathrm{t}}^{\mathrm{tr}}&=\sqrt{\tau p_{\mathrm{tr}} }\left( \bar{\bG}_{\mathrm{SR}}\bPhi_{\mathrm{S}}+{\bG}_{\mathrm{RD}}\bPhi_{\mathrm{D}} \right)+\bZ^{\mathrm{tr}}_{\mathrm{t}},
\end{align}
where  the channel matrices from the $K$ sources to the transmit antenna array of the relay station and from the $K$ destinations to the receive antenna array of the relay station are given by $\bar{\bG}_{\mathrm{SR}}\in \mathbb{C}^{M\times K}$ and $\bar{\bG}_{\mathrm{RD}}\in \mathbb{C}^{N\times K}$, respectively. Similarly, $\bZ^{\mathrm{tr}}_{\mathrm{r}}$ and $\bZ^{\mathrm{tr}}_{\mathrm{t}}$ denote AWGN matrices having i.i.d. $\mathcal{CN}\left( 0,1 \right)$ elements.  Also,  the $k$th
rows of $\bPhi_{\mathrm{S}}\in \mathbb{C}^{K\times \tau}$ and $\bPhi_{\mathrm{D}}\in \mathbb{C}^{K\times \tau}$ are the pilot sequences
transmitted from the corresponding source and destination users, i.e., $\bS_{k}$ and $\bD_{k}$. Actually, we assume that all the pilot sequences
are pairwisely orthogonal, which requires that $\tau \ge 2K$, since $\bPhi_{\mathrm{S}}\bPhi_{\mathrm{S}}^{\H}=\Id_{K}$, $\bPhi_{\mathrm{D}}\bPhi_{\mathrm{D}}^{\H}=\Id_{K}$, and $\bPhi_{\mathrm{S}}\bPhi_{\mathrm{D}}^{\H}=\b0$. Note that $p_{\mathrm{tr}}$ denotes the transmit power of each pilot symbol.

Under the assumption that the relay station applies minimum mean square-error (MMSE) estimation to estimate the channels $\bG_{\mathrm{SR}}$ and $\bG_{\mathrm{RD}}$, the estimated channels can be written by following the corresponding procedure in~\cite{Ngo2014}  as
\begin{align}
\hat{\bG}_{\mathrm{SR}}&=\frac{1}{\sqrt{\tau p_{\mathrm{tr}} }} \bY_{\mathrm{r}}^{\mathrm{tr}}\bPhi_{\mathrm{S}}^{\H}\tilde{\bD}_{\mathrm{SR}}\nn\\
&=\bG_{\mathrm{SR}}\tilde{\bD}_{\mathrm{SR}}+\frac{1}{\sqrt{\tau p_{\mathrm{tr}} }}\bN_{\mathrm{S}}\tilde{\bD}_{\mathrm{SR}}
\end{align}
and \begin{align}
\hat{\bG}_{\mathrm{RD}}&=\frac{1}{\sqrt{\tau p_{\mathrm{tr}}} } \bY_{\mathrm{t}}^{\mathrm{tr}}\bPhi_{\mathrm{D}}^{\H}\tilde{\bD}_{\mathrm{RD}}\nn\\
&=\bG_{\mathrm{RD}}\tilde{\bD}_{\mathrm{RD}}+\frac{1}{\sqrt{\tau p_{\mathrm{tr}} }}\bN_{\mathrm{D}}\tilde{\bD}_{\mathrm{RD}},
\end{align}
where $\bN_{\mathrm{S}}=\bZ^{\mathrm{tr}}_{\mathrm{t}}\bPhi_{\mathrm{D}}^{\H}$ and $\bN_{\mathrm{D}}=\bZ^{\mathrm{tr}}_{\mathrm{r}}\bPhi_{\mathrm{S}}^{\H}$. In addition, we have  $\tilde{\bD}_{\mathrm{SR}}=\left(  \tilde{\bD}_{\mathrm{SR}}^{-1}/\left( \tau p_{\mathrm{tr}} \right) +\Id_{K} \right)$ and  $\tilde{\bD}_{\mathrm{RD}}=\left(  \tilde{\bD}_{\mathrm{RD}}^{-1}/\left( \tau p_{\mathrm{tr}} \right) +\Id_{K} \right)$. Given that the  rows of $\bPhi_{\mathrm{S}}$ and $\bPhi_{\mathrm{D}}$ are pairwisely orthogonal, the
elements of $\bN_{\mathrm{S}}$ and $\bN_{\mathrm{D}}$ are  i.i.d. obeying to the $\mathcal{CN}\left( 0,1 \right)$ distribution.

Taking into account the property of orthogonality of MMSE estimation, we decompose the current channels in terms of the estimated channels as~\cite{Kay}
\begin{align}
{\bG}_{\mathrm{SR}}&=\hat{\bG}_{\mathrm{SR}}+{\bE}_{\mathrm{SR}}\\
{\bG}_{\mathrm{RD}}&=\hat{\bG}_{\mathrm{RD}}+{\bE}_{\mathrm{RD}},
\end{align}
where ${\bE}_{\mathrm{SR}}$ and ${\bE}_{\mathrm{RD}}$ are the estimation error matrices of ${\bG}_{\mathrm{SR}}$ and ${\bG}_{\mathrm{RD}}$. Actually, the rows of $\hat{\bG}_{\mathrm{SR}}$, ${\bE}_{\mathrm{SR}}$,  $\hat{\bG}_{\mathrm{RD}}$, and ${\bE}_{\mathrm{RD}}$ are mutually independent
and distributed as $\mathcal{CN}\left( \b0,\hat{\bD}_{\mathrm{SR}} \right)$,  $\mathcal{CN}\left( \b0,{\bD}_{\mathrm{SR}}-\hat{\bD}_{\mathrm{SR}} \right)$, $\mathcal{CN}\left( \b0,\hat{\bD}_{\mathrm{RD}} \right)$, and $\mathcal{CN}\left( \b0,{\bD}_{\mathrm{RD}}-\hat{\bD}_{\mathrm{RD}} \right)$. Note that $\hat{\bD}_{\mathrm{SR}}$ and $\hat{\bD}_{\mathrm{RD}}$  are diagonal matrices with $\left[ \hat{\bD}_{\mathrm{SR}}\right]_{kk}=\sigma^{2}_{\mathrm{SR},k}$ and $\left[ \hat{\bD}_{\mathrm{RD}}\right]_{kk}=\sigma^{2}_{\mathrm{RD},k}$  being the diagonal elements of $\hat{\bD}_{\mathrm{SR}}$, and $\hat{\bD}_{\mathrm{RD}}$, which are equal to $\sigma^{2}_{\mathrm{SR},k}=\tau p_{\mathrm{tr}}\beta^{2}_{\mathrm{SR},k}/\left( \tau p_{\mathrm{tr}}\beta_{\mathrm{SR},k}+1  \right)$ and $\sigma^{2}_{\mathrm{RD},k}=\tau p_{\mathrm{tr}}\beta^{2}_{\mathrm{RD},k}/\left( \tau p_{\mathrm{tr}}\beta_{\mathrm{RD,k}}+1  \right)$, respectively.

 \subsection{RS Approach}\label{RSApproach} 
After having described the  conventional multipair with relay transmission (NoRS) in Section~\ref{DataTransmission}, we focus on the application of the promising RS transmission method that is going to be applied in the second link between the relay station and the destination users. Below, we provide shortly its presentation. 

The main benefit of the RS transmission, taking place in multi-user scenarios, is the achievement of unsaturated sum-rate with increasing SNR despite the presence of imperfect CSIT as was shown in~\cite{Hao2015,Dai2016,Clerckx2016,Joudeh2016}. The NoRS strategy treats as noise every multi-user interference originating from the imperfect CSIT. On the other hand, the RS strategy is able to bridge treating interference as noise and perform interference decoding through the presence of a common message. Thus, the key to boost the sum-rate performance is the ability to decode part of the interference\footnote{ At the time of demodulation, a user needs to know the precoded channel to perform coherent demodulation. Actually, the user does not need to know the channel itself and the precoder itself, but just the inner product of the two, i.e. the precoded channel. The same action takes place  in conventional MU-MIMO. In practice, this is achieved through the use of demodulation reference signals, called DMRS in LTE-A~\cite{Lim2013}.}. This observation motivates us to investigate the potential benefits of RS  in the presence of the SI, since the SI has the effect of altering the CSI between the estimation stage and the transmission stage. 

According to the RS method, the message, intended for destination UE $k$, is split into two parts, namely, the common and  private parts. Regarding the  common part, it is drawn from a public codebook and it has to be decoded by all UEs with
zero error probability. As  far as the private part is concerned, it has to be decoded only by destination UE $k$. It is worthwhile  to mention that the messages, intended for the other UEs, consist of a private part only. 
In mathematical terms, the transmit signal is written as 
\begin{align}
 \bu &=\underbrace{\sqrt{ \rho_{\mathrm{c}}}\bff_{c} u_{c}}_{\mathrm{common ~part}}+\underbrace{\sum_{k=1}^{K}\sqrt{ \rho_{ \mathrm{k}}}\bff_{k} u_{k}}_{\mathrm{private~part}},\label{RStransmit}
\end{align}
where $u_{c}$  and $u_{k}$ are  the common  and the private messages for UE $k$, while $\bff_{c}$ denotes the  precoding vector of the common
message with unit norm and $\bff_{k}$ is the linear precoder corresponding to UE $k$. More concretely, the private message $u_{k}~ \forall k$ is superimposed over the common message $u_{c}$ and sent with linear precoding. In addition, $\rho_{\mathrm{c}}$ is the power allocated to the common message. Regarding the decoding procedure, the first step is the decoding of the common message  by each UE, while all private messages are treated as noise. The next step includes the subtraction of the contribution of the common message in the received signal by each UE, and thus, each UE is able to decode its own private message. {Herein, we focus on the application of the RZF precoder for the private messages, as mentioned before.} 
\begin{remark}[Conventional Transmission (NoRS Approach)]
According to the conventional approach, there is no common message transmission. Thus, since no common part exists, \eqref{RStransmit} degenerates to
\begin{align}
 \bu  =\sum_{k=1}^{K}{ \sqrt{\rho_{ \mathrm{k}}}}\bff_{k} u_{k},\label{NoRStransmit}
\end{align}
\end{remark}where $\lambda$ is a normalization parameter inside $\bff_{k}$ given by
$\lambda=\frac{K}{\EE \left[ \tr\bF^\H\bF \right]} $.
\section{End-to-End  Achievable Rate}\label{EndtoEndTransmission}
This section considers the presentation of the transmission between the $k$th source user and the corresponding destination user through the multiple antennas relay station, i.e., $\left( \mathrm{S}_{k}\to \mathrm{R}\to\mathrm{D}_{k} \right)$. Reasonably, this rate depends on the weakest link between the two hops, or else, this rate is limited by the minimum of the achievable rates of the two links~\cite{Riihonen2011}. More concretely, the achievable user rate from end-to-end is given by
\begin{align}
R_{k}=\min\{R_{\mathrm{SR},k},R_{\mathrm{RD},k} \},
\end{align}
where $R_{\mathrm{SR},k}$ and $R_{\mathrm{RD},k}$ denote the achievable rates of the corresponding links.

In the first transmission link, a conventional MAC is considered with an MMSE decoder at the relay station, while, in the second hop, we employ the RS   scheme with an RZF precoder for the transmission of the private messages.
\subsection{$\mathrm{S}_{k}\to \mathrm{R}$ (Conventional Transmission)}
During the first hop, we set $p_{\mathrm{S}}={\rho}$, where $\rho$ refers to the SNR, since the AWGN is assumed to have unit variance. Thus, the SINR  of the  source UE $k$ is expressed by means of~\eqref{BasicSystemModelwithoutRS4}  as 
  \begin{align}
   \gamma_{\mathrm{SR},k}=\frac{{\rho}|\bw_{k}^{\H}\bg_{\mathrm{SR},k}|^2}{\sum_{j\ne k}^{K}{\rho}|\bw_{k}^{\H}\bg_{\mathrm{SR},j}|^2 +\|\bw_{k}^{\H}\bG_{\mathrm{RR}}\|^2+\|\bw_{k}^{\H} \|^{2}}.\label{nors} 
\end{align}

Note that we have relied on the worst-case assumption by treating the multi-user interference and distortion noises as independent Gaussian noises~\cite{Hassibi2003}. According to this SINR, we obtain  the achievable sum-rate, being a lower bound of the mutual information between the received signal and the transmitted symbols, as  
\begin{align}
 \mathrm{R}_{\mathrm{SR}}&=\sum_{k=1}^{K}\mathrm{R}_{\mathrm{SR},k},\label{rate3} 
\end{align}
where $\mathrm{R}_{\mathrm{SR},k}=\frac{T-\tau}{T}\log_{2}\left( 1+ \gamma_{\mathrm{SR},k} \right)$. 

\subsection{$\mathrm{R}\to \mathrm{D}_{k}$}
During the second link, we employ the RS transmission scheme, in order to mitigate the saturation of the system at high SNR. Specifically, we apply uniform power allocation for  the private messages, however,  the power allocated to the common part is different. The allocation scheme assumes $\rho_{\mathrm{c}}=\rho\left( 1-t \right)$ to the common message and  $\rho_{\mathrm{k}}=\rho t/K$ to the private message of each UE, where $t \in \left( 0,1 \right]$. The $t$ parameter is used to adjust the  fraction of the total power spent on the transmission of the private messages.

Following the RS principles, we have to evaluate the SINRs of both common and private messages. Assuming that perfect CSI is available at the receivers and given that  the transmit  signal is given by~\eqref{RStransmit}, the corresponding SINRs are given  by
\begin{align}
   \gamma_{\mathrm{RD},k}^{\mathrm{c}}&=\frac{{\rho_{\mathrm{c}}}|\bg_{\mathrm{RD},k}^{\H}\bff_{\mathrm{c}}|^2}{\sum_{j=1}^{K}{\frac{\rho t}{K}|\bg_{\mathrm{RD},k}^{\H}\bff_{j}|^2 +1}}\label{c1} \\
  \gamma_{\mathrm{RD}}^{\mathrm{c}}&= \min_{k}\left( \gamma_{\mathrm{RD},k}^{\mathrm{c}} \right)\label{c2} \\
  \gamma_{\mathrm{RD},k}^{\mathrm{p}}&=\frac{\frac{\rho t}{K}|\bg_{\mathrm{RD},k}^{\H}\bff_{k}|^2}{\sum_{j\ne k}^{K}\frac{\rho t}{K}|\bg_{\mathrm{RD},k}^{\H}\bff_{j}|^2 +1}\label{c3}.
  \end{align}

Note that $\gamma_{\mathrm{RD}}^{\mathrm{c}} = \displaystyle  \min_{k}\left( \gamma_{\mathrm{RD},k}^{\mathrm{c}} \right)$ and $\gamma_{\mathrm{RD},k}^{\mathrm{p}}$ correspond to the SINRs of the common and private messages, respectively. In this case, the achievable sum-rate is written as
\begin{align}
 \mathrm{R}_{\mathrm{RD}}=\mathrm{R}^{\mathrm{c}}_{\mathrm{RD}}+\sum_{j=1}^{K}\mathrm{R}_{\mathrm{RD},j}^{\mathrm{p}},\label{RSSumRate} 
\end{align}
where, similar to~\eqref{rate3}, we have $\mathrm{R}^{\mathrm{c}}_{\mathrm{RD}}\!=\!\frac{T-\tau}{T}\log_{2}\!\left( 1\!+\! \gamma_{\mathrm{RD}}^{\mathrm{c}} \right)$  and $\mathrm{R}_{\mathrm{RD},k}^{\mathrm{p}}=\frac{T-\tau}{T}\log_{2}\left( 1+ \gamma_{\mathrm{RD},k}^{\mathrm{p}} \right)$  corresponding to the achievable rates of the common and private messages, respectively. 

\section{Deterministic Equivalent Performance Analysis}\label{Achievable}
The DEs of the  SINRs for both links are such that for each link it holds that  $\gamma_{k}-\overbar{\gamma}_{k}\xrightarrow[M \rightarrow \infty]{\mbox{a.s.}}0$\footnote{Note that $\xrightarrow[ M \rightarrow \infty]{\mbox{a.s.}}$ denotes almost sure convergence, and  $a_n\asymp b_n$ expresses the equivalence relation $a_n - b_n  \xrightarrow[ M \rightarrow \infty]{\mbox{a.s.}}  0$ with $a_n$  and $b_n$  being two infinite sequences.}, where $\gamma_{k}$ is the SINR of the $k$th user and $\bar{\gamma}_{k}$ is the corresponding DE. In this direction, the corresponding deterministic rate of UE $k$ is obtained by the dominated  convergence~\cite{Billingsley2008} and the continuous mapping theorem~\cite{Vaart2000} by means of~\eqref{rate3}, \eqref{RSSumRate}  for both links as
\begin{align}
R_{\mathrm{i},k}-\bar{R}_{\mathrm{i},k} \xrightarrow[ M \rightarrow \infty]{\mbox{a.s.}}0,~~~\mathrm{i}=\mathrm{SR}, \mathrm{RD}\label{DeterministicSumrate}
\end{align}
where  $\bar{R}_{\mathrm{i},k}$ is the DE $R_{\mathrm{i},k}$.

\subsection{DE of the  Achievable Rate of the First Hop  ($\mathrm{S}_{k}\to \mathrm{R}$)}
The design of the first hop, being basically a MAC, follows a standard uplink transmission scheme. We choose the MMSE linear decoder, in order to keep the implementation complexity to a reasonable level and at the same time achieve a high rate. 

The MMSE decoder is designed by means of the channel estimate $\hat{\bG}_{\mathrm{SR}}$, as~\cite{Hoydis2013}
\begin{align}
\bW_{\mathrm{SR}}
&= \left(\hat{\bW}_{\mathrm{SR}} \!+\!  \bZ_{\mathrm{SR}} \!+\! N \al_{\mathrm{RSR}} \Id_M\right)^{-1}{\hat{\bG}}_{\mathrm{SR}},
\end{align}
where we define
\begin{align}
 {\hat{\bSigma}}_{\mathrm{SR}}&\triangleq\left(\hat{\bW}_{\mathrm{SR}} \!+\!  \bZ_{\mathrm{SR}} \!+\! N \al_{\mathrm{SR}} \Id_M\right)^{-1}
\end{align}
with $\hat{\bW}_{\mathrm{SR}}\triangleq\hat{\bG}_{\mathrm{SR}}\hat{\bG}_{\mathrm{SR}}^{\H}$.  The matrix $\bZ_{\mathrm{SR}} \in \bbC^{N \times N}$ is an arbitrary Hermitian
nonnegative definite matrix  and $\al_{\mathrm{SR}}$ is a regularization parameter scaled by $N$, in order to converge to a constant, as $N$, $K\to \infty$. Although $\al_{\mathrm{RD}}$ and $\bZ_{\mathrm{RD}}$ can be optimized, this is outside the scope of this paper and we leave it for future work. 

The data transmission during this hop has a duration of $T-\tau$ time slots. The DE of the $k$ user rate, when $K,~N$ go to infinity with a given ration $\beta=N/K$, is provided by the following theorem.

\begin{Theorem}\label{theorem:MMSE}
The  DE of the SINR of UE $k$  for the first link of a  multipair FD system with MMSE  decoding  and imperfect CSIT is given by~\eqref{privateLINR}, 
\begin{figure*}
\begin{align}
\overbar{\gamma}_{\mathrm{SR},k}&=\frac{{\rho}{\delta}_{\mathrm{SR},k} ^{2}}{{\rho}\frac{1}{N}{\delta}_{\mathrm{SR},k}^{'}+\frac{1}{N}{\delta}_{\mathrm{SR},k}^{''}+
 \frac{1}{N} \tr \bT_{\mathrm{RR}}+{\rho}\sum_{j\ne k}^{K}\frac{{\mu}_{\mathrm{SR},jk}}{N}}\label{privateLINR} 
\end{align}
\line(1,0){470}
\end{figure*}
where
   \begin{align}
  {\mu}_{jk} \asymp  \frac{ \delta_{j}^{'}}{N}  + \frac{\left|{\delta_{k}^{''}}\right|^{2}\delta_{k}^{'}}{N\left( 1 + \delta_{k} \right)^{2}} - 2\mathrm{Re}\left\{   \frac{ {\delta}_{k}^{''}\delta_{k}^{'} }{N\left( 1 + \delta_{k} \right)} \right\}\!.
\end{align}
Also, we have 
${\delta}_{\mathrm{SR},k}=\frac{1}{N}\tr \hat{\bD}_{\mathrm{SR},k}\bT_{\mathrm{SR},k}$,
${\delta}_{\mathrm{SR},k}^{'}=\frac{1}{N}\tr \hat{\bD}_{\mathrm{SR},k}\bT_{\mathrm{SR},k}^{'}$, 
${\delta}_{\mathrm{SR},k}^{''}=\frac{1}{N}\tr \hat{\bD}_{\mathrm{SR},k}\bT_{\mathrm{SR},k}^{''}$, 
$\delta_{\mathrm{SR},jk}=\frac{1}{N}\tr \hat{\bD}_{\mathrm{SR},k}{\bT}_{\mathrm{SR},jk}$, $\delta_{\mathrm{SR},jk}^{'}=\frac{1}{N}\tr \hat{\bD}_{\mathrm{SR},j}{\bT}_{\mathrm{SR},jk}^{'}$, $\delta_{jk}^{''}=\frac{1}{N}\tr \hat{\bD}_{\mathrm{SR},jk}{\bT}_{\mathrm{SR},jk}$, ${\bS}=\bZ_{\mathrm{SR}} /N$, and $\tilde{a}=\al_{\mathrm{SR}}  $
 where
\begin{itemize}
\renewcommand{\labelitemi}{$\ast$}
\item $\bT_{\mathrm{SR},k}=\bT_{\mathrm{SR},k}(\tilde{a})$ and ${\deltav}=[{\delta}_{1},\cdots,{\delta}_{K}]^\T={\deltav}(\tilde{a})={\ev}(\tilde{a})$ are given by~\cite[Thm. 1]{Wagner2012} for ${\bS}=\bS$, $\bD=\hat{\bD}_{\mathrm{SR},k},~\bL=\hat{\bD}_{\mathrm{SR},k}\, \forall k \in \mathcal{K}$,
\item  $\bT_{\mathrm{SR},k}=\bT_{\mathrm{SR},k}^{'}(\tilde{a})$ is given by~\cite[Thm. 2]{Hoydis2013} for  ${\bS}=\bS$, $\bD=\hat{\bD}_{\mathrm{SR},k}$, $\bK=\bD_{\mathrm{SR},k} - \hat{\bD}_{\mathrm{SR},k}, \forall  k \in \mathcal{K}$,
\item   $\bT_{\mathrm{SR},k}=\bT_{\mathrm{SR},k}^{''}(\tilde{a})$ is given by~\cite[Thm. 2]{Hoydis2013} for ${\bS}=\bS$, $\bD= \hat{\bD}_{\mathrm{SR},k}$, $\bK= \Id_{N}, \forall  k \in \mathcal{K}$,
\item  $\bT_{\mathrm{RR}}=\bT_{\mathrm{RR}}(\tilde{a})$ is given by~\cite[Thm. 2]{Hoydis2013} for  ${\bS}=\bS$, $\bD=\Id_{N}$, $\bK=\sigma^{2}_{\mathrm{SI}}\Id_{N}, \forall  k \in \mathcal{K}$,
\item  $\bT_{\mathrm{SR},jk}=\bT_{\mathrm{SR},jk}(\tilde{a})$ is given by~\cite[Thm. 2]{Hoydis2013} for  ${\bS}=\bS$, $\bD=\hat{\bD}_{\mathrm{SR},k}$, $\bK=\hat{\bD}_{\mathrm{SR},k}, \forall  k \in \mathcal{K}$,
\item  $\bT_{\mathrm{SR},jk}=\bT_{\mathrm{SR},jk}^{'}(\tilde{a})$ is given by~\cite[Thm. 2]{Hoydis2013} for ${\bS}=\bS$, $\bD= \hat{\bD}_{\mathrm{SR},k}$, $\bK= \hat{\bD}_{\mathrm{SR},k},~\bL= \hat{\bD}_{\mathrm{SR},j} \forall  k \in \mathcal{K}$.
\item $\bT_{\mathrm{SR},jk}=\bT_{\mathrm{SR},jk}^{''}(\tilde{a})$ is given by~\cite[Thm. 2]{Hoydis2013} for ${\bS}=\bS$, $\bD= \hat{\bD}_{\mathrm{SR},k}$, $\bK= \hat{\bD}_{\mathrm{SR},k},~\bL= \hat{\bD}_{\mathrm{SR},k} \forall  k \in \mathcal{K}$. 
\end{itemize} 
\end{Theorem}
\proof The proof of Theorem~\ref{theorem:MMSE} is given in Appendix~\ref{theorem4}.\endproof

\subsection{DE of the  Achievable Rate of the Second Hop with RS ($\mathrm{R}\to \mathrm{D}_{k}$)}
This section  presents the DE of the user rate during the data transmission with RS in the second link,  which takes place for $T-\tau$ time slots. In fact, we derive the DE of the $k$th UE in the asymptotic limit of $K, M$ for fixed ratio $\zeta=K/M$.

In addition, we provide the precoder design for the common message, implemented to be used under the RS approach. Moreover, among the main results, we present the DEs of the SINRs characterizing the transmissions of the common  and the private messages of UE $k$.

\subsection{Precoder Design}\label{PD} 
For the sake of simplicity, we employ linear precoding during the application of the RS method. Actually, the RS method includes two different types of precoders for the transmission of the private and common messages, respectively. In the case of a MISO broadcast channel (BC) with imperfect CSI, the optimal precoder has to be optimized numerically~\cite{Joudeh2016}. However, we consider that the transmission of the private message takes place by using RZF due to the prohibitive complexity, as mentioned in a previous section\footnote{Note that an extra  gain of RS over NoRS can be achieved by jointly optimizing the power allocation as well as the precoders of the common and private messages~\cite{Joudeh2016}. However it is not really practical to resort to this type of optimization for large-scale systems such as massive MIMO, where the use of deterministic equivalent analysis is commonly used in order to get some further insight into the system behaviour in terms of different aspects such as the impact of hardware impairments~\cite{Papazafeiropoulos2017}.}.  Further elaboration follows.
\subsubsection{Precoding of the Private Messages}
Given that the  complexity increases in large MIMO systems as $M \to \infty$, the choice of RZF for the transmission of the private messages is the prevailing solution.  In such case, the relay station  implements its RZF precoder, constructed by means of the channel estimate $\hat{\bG}_{\mathrm{RD}}$, as~\cite{Hoydis2013}
\begin{align}
\bF_{\mathrm{RD}}
&= \sqrt{\lambda}\left(\hat{\bW}_{\mathrm{RD}} \!+\!  \bZ_{\mathrm{RD}} \!+\! M \al_{\mathrm{RD}} \Id_M\right)^{-1}{\hat{\bG}}_{\mathrm{RD}}\nn\\
&= \sqrt{\lambda} {\hat{\bSigma}}_{\mathrm{RD}} {\hat{\bG}}_{\mathrm{RD}}, \label{eq:precoderRZF}
\end{align}
where we define
\begin{align}
 {\hat{\bSigma}}_{\mathrm{RD}}&\triangleq\left(\hat{\bW}_{\mathrm{RD}} \!+\!  \bZ_{\mathrm{RD}} \!+\! M \al_{\mathrm{RD}} \Id_M\right)^{-1}
\end{align}
with $\hat{\bW}_{\mathrm{RD}}\triangleq\hat{\bG}_{\mathrm{RD}}\hat{\bG}_{\mathrm{RD}}^{\H}$ and $\lambda$ being a normalization parameter that satisfies
$\lambda=\frac{K}{\EE \left[ \tr\bF_{\mathrm{RD}}^\H\bF_{\mathrm{RD}}\right]}$,
which is a long-term total transmit power constraint at the relay.  Similar to the definition of the MMSE decoder,  $\bZ_{\mathrm{RD}} \in \bbC^{M \times M}$ is an arbitrary Hermitian
nonnegative definite matrix  and $\al_{\mathrm{RD}}$ is a regularization parameter scaled by $M$, in order to converge to a constant, as $M$, $K\to \infty$. In addition, $\al_{\mathrm{RD}}$ and $\bZ_{\mathrm{RD}}$ can be optimized as well, but this is outside the scope of this paper and we leave it for future work. 
\subsubsection{Precoding of the Common Message}
Herein, we provide the design of the precoder $\bff_{c}$ of the common message by following a similar procedure to~\cite{Dai2016}. In particular, taking into account that in the large number of antennas regime the  different channel estimates tend to be orthogonal, we express   $\bff_{c}$ as a linear sum of these channel estimates in the subspace of $\hat{\bG}_{\mathrm{RD}}$,  $\mathcal{S}=\mathrm{Span}\left( \hat{\bG}_{\mathrm{RD}} \right)$. In other words, $\bff_{c}$ is designed as a weighted matched beamforming. Mathematically, we have
\begin{align}
 \bff_{c}=\sum_{k}\alpha_{k} \hat{\bg}_{ \mathrm{RD},k}.
\end{align}

The objective is to maximize the achievable
rate of the common message $\mathrm{R}_{ \mathrm{RD},k}^{\mathrm{c}}$. This optimization problem is formed as
\begin{align}\begin{split}
&\mathcal{P}_{1}~:~\max_{\bff_{c} \in \mathcal{S}}\,\min_{k} {q}_{ \mathrm{RD},k}|\bg_{\mathrm{RD},k}^{\H} \bff_{c}|^2,\\
 &\mathrm{s.t.}~~~~\|\bff_{c}\|^{2}=1\label{P1} 
\end{split}
\end{align}
where $q_{ \mathrm{RD},k}=\frac{{\rho_{\mathrm{c}}}\lambda}{\lambda\sum_{j=1}^{K}\frac{\rho}{K}|\bg_{ \mathrm{RD},k}^{\H} \bff_{j}|^2+1}$. The optimal  $\{\al_{k}^{*}\}$ is yielded by  the following proposition.

\begin{proposition}
In the large system limit, the optimal solution of the practical problem set by $\mathcal{P}_{1}$  is given by
\begin{align}
 \al^{*}_{k}=\frac{1}{\sqrt{M \sum_{j=1}^{K}\frac{q_{k}\frac{1}{M^{2}}\tr^{2}\hat{\bD}_{\mathrm{RD},k}}{q_{j} \frac{1}{M^{2}}\tr^{2}\hat{\bD}_{\mathrm{RD},j}}}},~\forall k.
\end{align}
\end{proposition}
\proof  We achieve to result in   an optimization problem with deterministic variables  by obtaining the DEs of the equation and the constraint of $\mathcal{P}_{1}$. Indeed, applying  Lemma~\cite[Lem. B.26]{Bai2010a} to \eqref{P1}, we have 
 \begin{align}\begin{split}
&\mathcal{P}_{2}~:~\max_{\al_{k} }\,\min_{k} {q}_{ \mathrm{RD},k}\frac{1}{M^{2}} |\al_{k}    \tr\hat{\bD}_{\mathrm{RD},k}\
|^{2},\\
 &\mathrm{s.t.}~~~~\sum_{k}\al_{k}^{2}=\frac{1}{M}.\label{p2} 
\end{split}
\end{align}

Use of Lemma $2$ in~\cite{Xiang2014} indicates that the optimal solution, satisfying $\mathcal{P}_{2}$, results, if   all terms are equal. Specifically, the optimal solution is found when $q_{ \mathrm{RD},k}\al_{k}^{2} \frac{1}{M^{2}}
                          \tr^{2}\hat{\bD}_{\mathrm{RD},k}=q_{ \mathrm{RD},j}\al_{j}^{2} \frac{1}{M^{2}}\tr^{2}\hat{\bD}_{\mathrm{RD},j},
                       $ $\forall k\ne j$, and the proof is concluded.
\endproof

\begin{Theorem}\label{theorem:RZF}
The  DEs of the SINRs of UE $k$ for the second link of a  multipair FD system, corresponding to the  private and common messages with RZF precoding  and imperfect CSIT, are given by
\begin{align}
 \overbar{\gamma}_{\mathrm{RD},k}^{\mathrm{p}}&=\frac{\bar{\lambda}\frac{\rho t}{K}{\delta}_{\mathrm{RD},k}^{2}}{\bar{\lambda}\frac{\rho t}{K}\displaystyle \sum_{j\ne k}^{K}\frac{{Q}_{\mathrm{RD},jk}}{M}+\left(1+{\delta_{\mathrm{RD},k}}\right)^{2}}\label{privateLINR1} \\
\overbar{\gamma}_{\mathrm{RD},k}^{c}&=\frac{\rho_{\mathrm{c}}\bar{\lambda}\alpha_{k}^{2} {\delta}_{\mathrm{RD},k}^{2}}{\bar{\lambda}\frac{\rho t}{K}\displaystyle\sum_{j=1}^{K}\frac{{Q}_{\mathrm{RD},jk}}{M}+\left(1+{\delta_{\mathrm{RD},k}}\right)^{2}},\label{CommonLINR}
 \end{align}
 where 
  \begin{align}
  \bar{\lambda}&=K\left( \frac{1}{M}\sum_{k=1}^{K}\frac{{\delta}_{\mathrm{RD},k}^{'}}{\left( 1+{\delta}_{\mathrm{RD},k} \right)^{2}} \right)^{-1},\nn
  \end{align}
  and
  \begin{align}
  {Q}_{\mathrm{RD},jk} \asymp  \frac{ \delta_{\mathrm{RD},jk}^{''}}{M}  &+ \frac{\left|{\delta_{\mathrm{RD},jk}^{'''}}\right|^{2}\delta_{\mathrm{RD},jk}^{''}}{M\left( 1 + \delta_{\mathrm{RD},k} \right)^{2}} \nn\\
&- 2\mathrm{Re}\left\{   \frac{ {\delta}_{\mathrm{RD},jk}^{'''}\delta_{\mathrm{RD},jk}^{''} }{M\left( 1 + \delta_{\mathrm{RD},k} \right)} \right\} .
 \label{eq:theorem2.I.mu}
\end{align}
Also, we have 
${\delta}_{\mathrm{RD},k}=\frac{1}{M}\tr \hat{\bD}_{\mathrm{RD},k}\bT_{\mathrm{SR},k}$, $\delta_{\mathrm{RD},jk}^{'}=\frac{1}{M}\tr \hat{\bD}_{\mathrm{RD},k}{\bT}^{'}_{\mathrm{SR},k}$, $\delta_{\mathrm{RD},jk}^{'}=\frac{1}{M}\tr \hat{\bD}_{\mathrm{RD},j}{\bT}^{'}_{\mathrm{SR},k}$, $\delta_{\mathrm{RD},jk}^{''}=\frac{1}{M}\tr \hat{\bD}_{\mathrm{RD},jk}{\bT}^{''}_{\mathrm{SR},k}$, $\delta_{\mathrm{RD},jk}^{'''}=\frac{1}{M}\tr \hat{\bD}_{\mathrm{RD},jk}{\bT}^{'''}_{\mathrm{SR},k}$, ${\bS}=\bZ_{\mathrm{RD}} /M$, and $\tilde{a}=\al_{\mathrm{RD}} $
 where
\begin{itemize}
\renewcommand{\labelitemi}{$\ast$}
\item $\bT_{\mathrm{SR},k}=\bT_{\mathrm{SR},k}(\tilde{a})$ and ${\deltav}=[{\delta}_{1},\cdots,{\delta}_{K}]^\T={\deltav}(\tilde{a})={\ev}(\tilde{a})$ are given by~\cite[Thm. 1]{Wagner2012} for ${\bS}=\bS$, $\bD_{\mathrm{SR},k}=\hat{\bD}_{\mathrm{SR},k}\, \forall k \in \mathcal{K}$,
\item $\bT_{\mathrm{SR},jk}=\bT_{\mathrm{SR},jk}^{}(\tilde{a})$ is given by~\cite[Thm. 1]{Wagner2012} for ${\bS}=\bS$,   $\bD_{\mathrm{SR},k}=\hat{\bD}_{\mathrm{SR},k}\, \forall k \in \mathcal{K}$,
\item $\bT_{\mathrm{SR},jk}=\bT_{\mathrm{SR},jk}^{'}(\tilde{a})$ is given by~\cite[Thm. 2]{Hoydis2013}  for ${\bS}=\bS$, $\bL_{\mathrm{SR},k}=\Id_{M}$,  $\bD_{\mathrm{SR},k}=\hat{\bD}_{\mathrm{SR},k}\, \forall k \in \mathcal{K}$,
\item  $\bT_{\mathrm{SR},jk}=\bT_{\mathrm{SR},jk}^{''}(\tilde{a})$ is given by~\cite[Thm. 2]{Hoydis2013} for  ${\bS}=\bS$, $\bK= \hat{\bD}_{\mathrm{SR},j}$, $\bK=\hat{\bD}_{\mathrm{SR},j}$, $\bD_k=\hat{\bD}_{\mathrm{SR},k}, \forall  k \in \mathcal{K}$,
\item  $\bT_{\mathrm{SR},jk}=\bT_{\mathrm{SR},jk}^{'''}(\tilde{a})$ is given by~\cite[Thm. 2]{Hoydis2013} for ${\bS}=\bS$, $\bL= \hat{\bD}_{\mathrm{SR},k}$ $\bK= \hat{\bD}_{\mathrm{SR},k}$, $\bD_k= \hat{\bD}_{\mathrm{SR},k}, \forall  k \in \mathcal{K}$.
\end{itemize} 
\end{Theorem}
\proof The proof of Theorem~\ref{theorem:RZF} is given in Appendix~\ref{theorem3}.\endproof

The following remark will enable us to shed light on the interesting properties of multipair FD systems with a large number of relay station antennas during the presentation and investigation of the numerical results.
\begin{remark}[Impact of increasing transmit and receive relay station antennas]
According to~\cite{Ngo2014}, the impact of SI cancels out, when the SI is projected onto its orthogonal complement. Unfortunately, following this direction, the  orthogonal projection may harm the desired signal, unless the receive antenna array grows large (tending to infinity). In such case, the channel vectors of the desired signal and the loop interference become nearly orthogonal, and actually, the impact of SI is reduced.

In a parallel path, if the size of the transmit antenna array is increased, the relay station will be able to  focus its emitted energy into the proper destination users. Moreover, the transmission towards the receive antennas  of the relay station is avoided. Hence, the SI reduces almost to zero.
\end{remark}
\begin{remark}[Reduction to HD transmission]
Changing the system model, describing the FD transmission, to HD transmission by neglecting the SI term and changing the prelog factor in the achievable rate, we reduce to the expressions providing the DE rates of the private and common messages corresponding to \eqref{privateLINR1} and \eqref{CommonLINR}.
\end{remark}

\subsection{Power Allocation}
The normal method to obtain the optimal power splitting ratio $t$,  maximizing~\eqref{RSSumRate}, includes the derivation of the first-order derivative. However, the complicated form of the solution, led us to follow a suboptimal power allocation
method similar to~\cite{Dai2016}, where RS outperforms the conventional broadcasting schemes. Interestingly, the solution allows us to extract useful observations.  According to the main idea, the allocation of the fraction $t$ results by setting  the
total transmit power of  the private messages of RS, in order to achieve approximately the same sum rate as the conventional multi-user BC with full power. The remaining power is allocated for the transmission of the common message of RS, which  boosts the sum rate. 
The gain in the sum-rate of the second link, achieved by the RS strategy with comparison to the NoRS transmission, is given by the difference
\begin{align}
 \Delta R_{\mathrm{RD}}=\mathrm{
 R}^{\mathrm{c}}_{\mathrm{RD}}+\sum_{k=1}^{K}\left( \mathrm{R}_{{\mathrm{RD},k}}^{\mathrm{p}}-\mathrm{R}_{{\mathrm{RD},k}}^{\mathrm{NoRS}} \right)\label{DR} .
\end{align}

\begin{proposition}\label{prop:inequality} 
The necessary condition, described by $\mathrm{R}_{{\mathrm{RD},k}}^{\mathrm{p}}\le\mathrm{R}_{{\mathrm{RD},k}}^{\mathrm{NoRS}}$, becomes equality, when the power splitting ratio $t$ is given by
\begin{align}
\!\!\!t=\min\bigg\{\frac{K}{\rho \bar{Y}},1\bigg\},\label{tau} 
\end{align}
where $\bar{Y}=\frac{\bar{\lambda}\frac{1}{K}\left(   {\delta}_{\mathrm{RD},k} \right)^{2}}{\bar{\lambda}\frac{ 1}{K}\displaystyle \sum_{j\ne k}^{K}\frac{{Q}_{\mathrm{RD},jk}}{M}+\left(1+ {\delta_{\mathrm{RD},k}}\right)^{2}}$.
In such case, the sum-rate gain $\Delta R_{\mathrm{RD},k}$ becomes
\begin{align}
 \Delta R_{\mathrm{RD}}\ge \mathrm{R}_{\mathrm{RD}}^{\mathrm{c}}-\log_2 e.
\end{align}
\end{proposition}
\proof See Appendix~\ref{proofinequality}.\endproof

\section{Numerical Results}\label{NumericalResults} 
This section presents the verification of the accuracy of the derived DE expressions (analytical results) by means of comparison with the Monte Carlo simulation results. Moreover, the numerical illustrations allow to gain  insights on the system performance of the considered model, and mostly on the impact of SI. In particular, the bullets represent the simulation results. 
\subsection{Simulation Setup}
We consider a Rayleigh block-fading channel, where the coherence time and the coherence bandwidth are $T_{c}=5~\mathrm{ms}$ and $B_{c}=100~\mathrm{KHz}$, respectively. As a result, the coherence block consists of $T = 500$ channel uses.
The simulation topology assumes  $K = 10$ communication pairs, located randomly inside a disk with a diameter of $1000$ $\mathrm{m}$. The pilot length is $\tau = 20$.  In each block, we assume fast fading by means of $\bh_{\mathrm{SR},k} \sim \cC\cN(\b0,\Id_{N})$ and $\bh_{\mathrm{RD},k} \sim \cC\cN(\b0,\Id_{K})$. Also,  we account for path-loss and shadowing, where $\bD_{\mathrm{RS}}$ is a $K \times K$ diagonal matrix with elements across the diagonal modeled as
 $\beta_{\mathrm{SR},k}^{m}=\frac{10^{s_{\mathrm{SR},k}^{-1.53}}}{d_{\mathrm{SR},k} ^{3.76}}$ with $d_{\mathrm{SR},k}$  being the distance in meters between the  receive antenna $m$ at the relay station and source UE $k$, and $s_{\mathrm{SR},k}\sim \cN\left(0,3.16\right)$ representing the shadowing effect~\cite{Bjornson2015}. Without loss of generality, we assume the same large-scale conditions for the second link. In addition, the  power of the uplink training symbols for both links is  $p_{\mathrm{tr}}=2~\mathrm{dB}$~\cite{Hoydis2013}.
The number of transmit and receive antennas is $M=N=100$. Note that these parameters hold throughout this section, unless otherwise stated.

\subsection{Robustness of RS in FD systems?-Comparisons}
The metric, we employ, to shed light on this meaningful question is the theoretical DE sum-rate  and the corresponding Monte Carlo simulation. Actually, the theoretical curves are obtained by means of Theorems~\ref{theorem:MMSE} and \ref{theorem:RZF} as well as \eqref{rate3} and \eqref{RSSumRate} by means of \eqref{DeterministicSumrate}. On the other hand, the simulated curves are obtained by averaging the corresponding rate over $10^{3}$ random channel instances. The choice of $t$ took place by means of Proposition~\ref{prop:inequality}. Although the DEs are derived for $K$, $M$, and $N\to \infty$ with  given ratios, they coincide with the simulated curves even for finite values $K$, $M$, and $N$\footnote{It is shown that the simulations  coincide with the DEs for $M=N=20$ number of antennas. Therefore, the DEs provide reliable results even for low system dimensions. It should be noted that this is not a new observation. Similar observations have been made in the literature even for an $8 \times 8$ system~\cite{Hoydis2013,Vaart2000,Wagner2012,Bai2010a}.}.
\begin{figure}[!h]
 \begin{center}
     \includegraphics[width=0.95\linewidth]{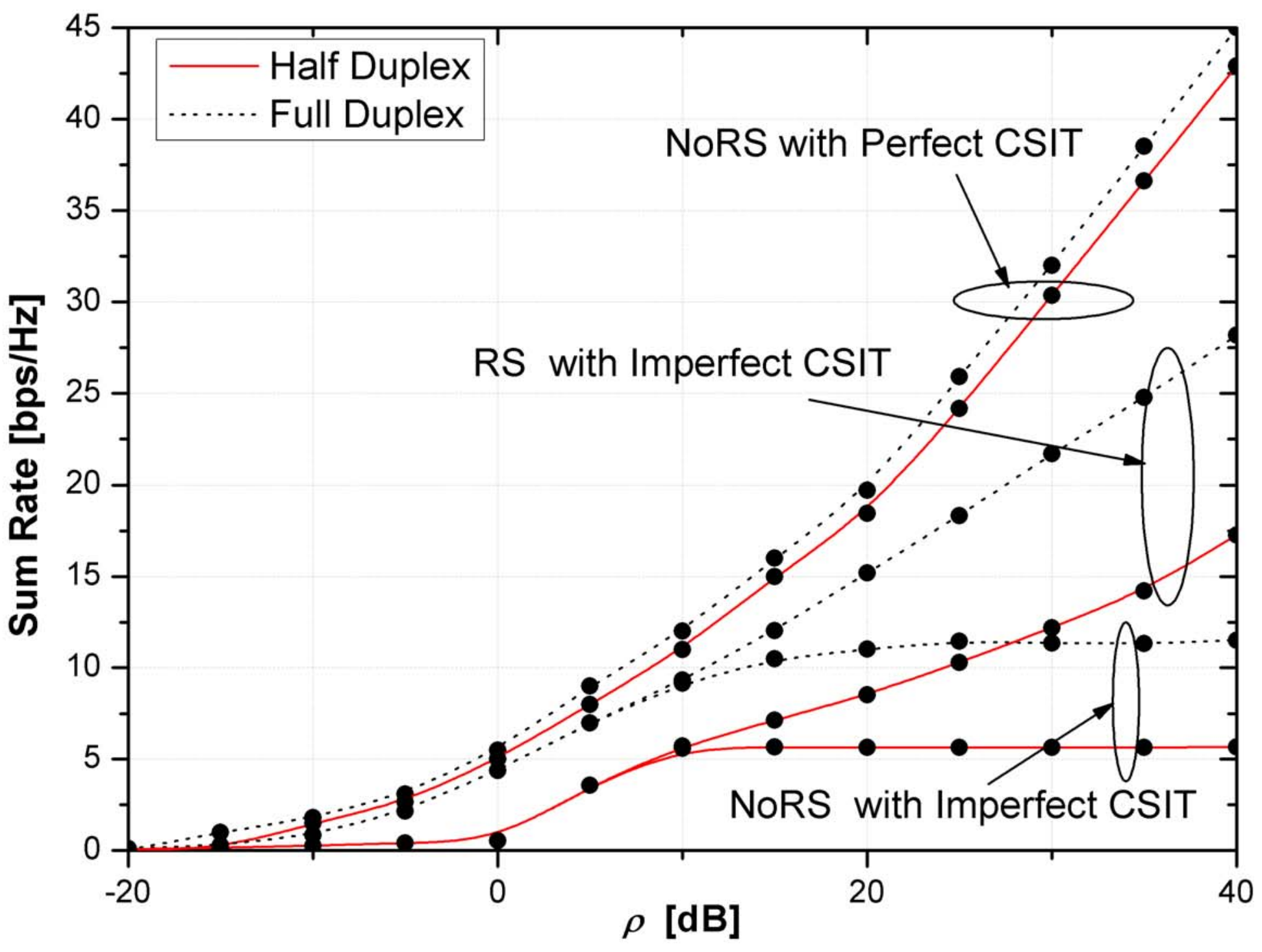}
 \caption{\footnotesize{Sum-rate versus $\rho$ for different transmission techniques and comparison between HD and FD ($M=100$, $K=10$, $T=500$, $p_{\mathrm{tr}}=2~\mathrm{dB}$, $\sigma^{2}_{\mathrm{SI}}=0$ $\mathrm{dB}$).}} 
 \label{FH_HD}
 \end{center}
 \end{figure}
 
\subsubsection{Comparison between FD and HD} Fig.~\ref{FH_HD} provides the comparison between FD and HD strategies in different transmission settings, when  $\sigma^{2}_{\mathrm{SI}}=0$ $\mathrm{dB}$. Specifically, the dashed black lines denote the FD method, while the red solid lines depict the HD method. In all cases, FD outperforms HD as expected. In addition, we show that RS with perfect CSIT for both FD and HD increase with $\rho$ without bound. The practical scenario, where CSIT is imperfect, is depicted by means of rate saturation in the case of NoRS, while RS provides an unsaturated rate. Although the gap between FD and HD is kept constant at high SNR in the case of NoRS, RS appears to be even more preferable at the same SNR regime, as expected. Hence, RS proves to be  robust since the rate does not saturate, and it is even more preferable at high SNR. Overall, RS is beneficial for both FD and HD, but FD outperforms HD due to the prelog factor and the mitigation of SI, especially at high SNR (increasing gap with increasing $\rho$).

\subsubsection{Varying the severity of SI} 
\begin{figure}[!h]
 \begin{center}
 \includegraphics[width=0.95\linewidth]{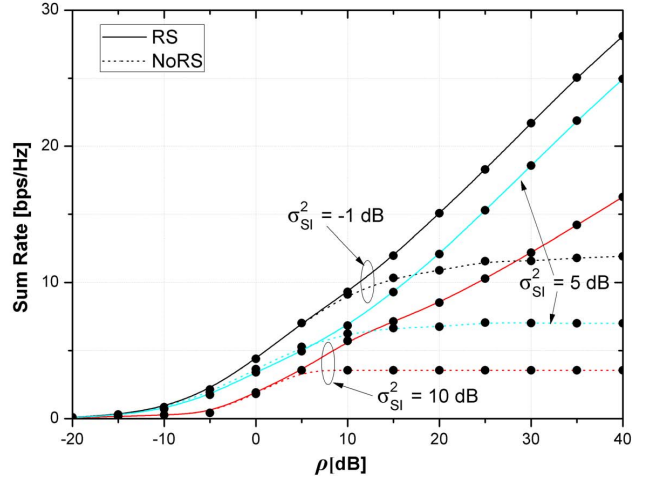}
 \caption{\footnotesize{Sum-rate versus $\rho$  for  varying $\sigma^{2}_{\mathrm{SI}}$ ($M=100$, $K=10$, $T=500$, $p_{\mathrm{tr}}=2~\mathrm{dB}$).}}
 \label{VaryingSigma2}
 \end{center}
 \end{figure}

Fig.~\ref{VaryingSigma2} presents the sum-rate versus the SNR for varying $\sigma^{2}_{\mathrm{SI}}$. Actually, the solid lines correspond to RS, while the dashed lines represent the implementation of NoRS. The highest line corresponds to the smallest value of $\sigma^{2}_{\mathrm{SI}}$, i.e., $\sigma^{2}_{\mathrm{SI}}=-1$ $\mathrm{dB}$, because in FD, the least the $\sigma^{2}_{\mathrm{SI}}$ is, the largest the rate becomes. When $\sigma^{2}_{\mathrm{SI}}=10$ $\mathrm{dB}$, the slope is less than the other cases showing that the higher the $\sigma^{2}_{\mathrm{SI}}$ is, the less capable RS is of mitigating the SI.
 
\begin{figure}[!h]
 \begin{center}
 \includegraphics[width=0.95\linewidth]{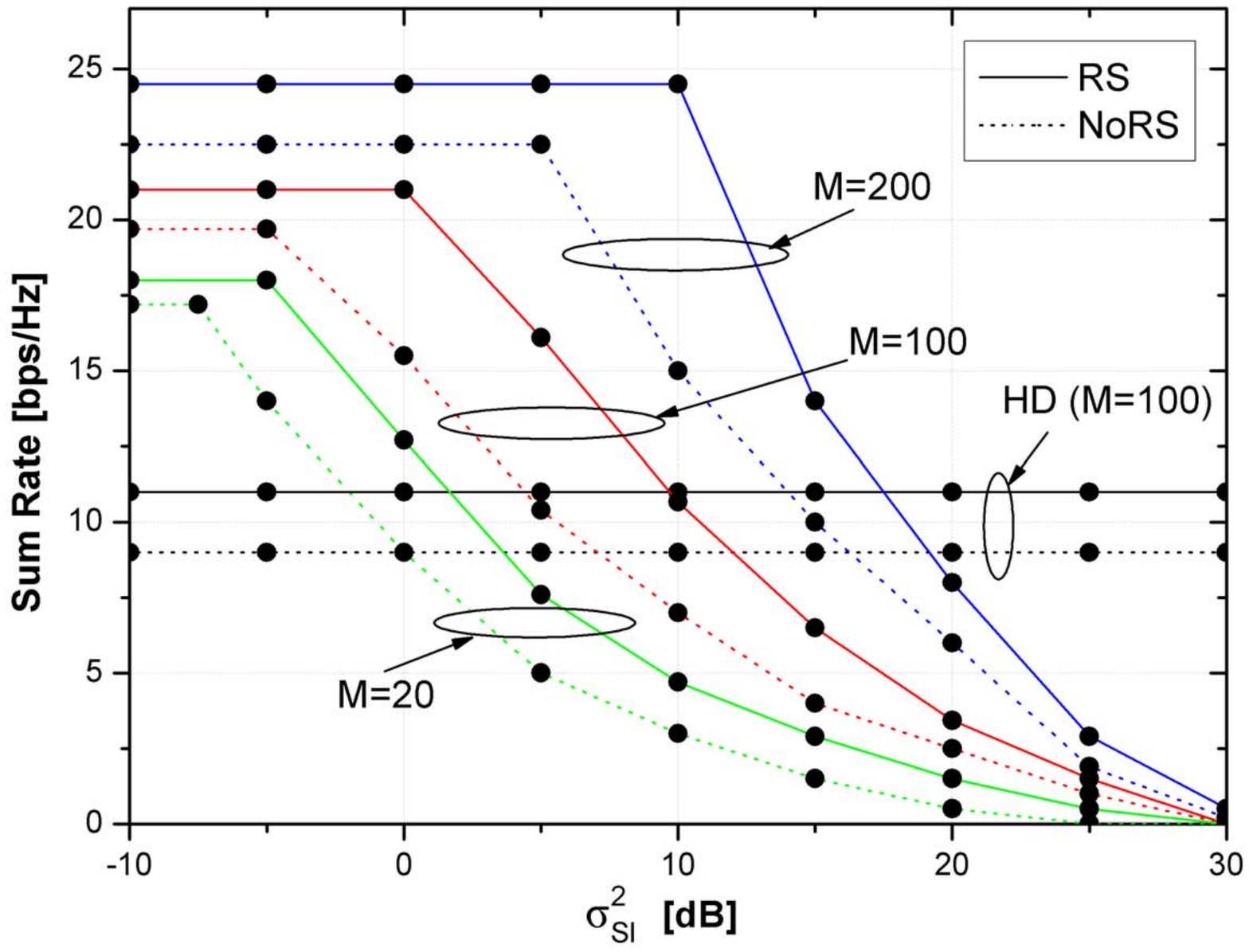}
 \caption{\footnotesize{Sum-rate versus $\sigma^{2}_{\mathrm{SI}}$ for varying $M$ ($\sigma^{2}_{\mathrm{SI}}=0$ $\mathrm{dB}$, $K=10$, $T=500$, $p_{\mathrm{tr}}=2~\mathrm{dB}$).}}
 \label{VaryingSigmaHorizontally}
 \end{center}
 \end{figure}

Fig.~\ref{VaryingSigmaHorizontally} illustrates the sum-rate versus the variance of the SI for $\rho=20$ dB, when the number of transmit antennas of the relay station $M$ increases. At the same figure, we have plotted the sum-rate corresponding to the HD case, which does not depend on~$\sigma^{2}_{\mathrm{SI}}$, i.e., it is constant with regard to $M$ (parallel to the horizontal axis). Furthermore, it is exposed that, as the number $M$ increases, the impact of SI becomes less severe, and hence, the sum-rate is higher. Until a specific value of $\sigma^{2}_{\mathrm{SI}}$, the sum-rate does not change, which means that the impact of SI is negligible. After this value, the sum-rate starts decreasing. Moreover, in the case that $M=100$, we can observe that HD appears with higher sum-rate, if the impact from $ \sigma^{2}_{\mathrm{SI}}$ is large enough, i.e., $\sigma^{2}_{\mathrm{SI}}>18$ $\mathrm{dB}$. For the sake of comparison, we have included the plots corresponding to the NoRS transmission. It is revealed that RS provides an increase in the range of SI over which FD outperforms HD.

\subsubsection{Varying the Number of Relay Station Transmit Antennas $M$} 
\begin{figure}[!h]
 \begin{center}
 \includegraphics[width=0.95\linewidth]{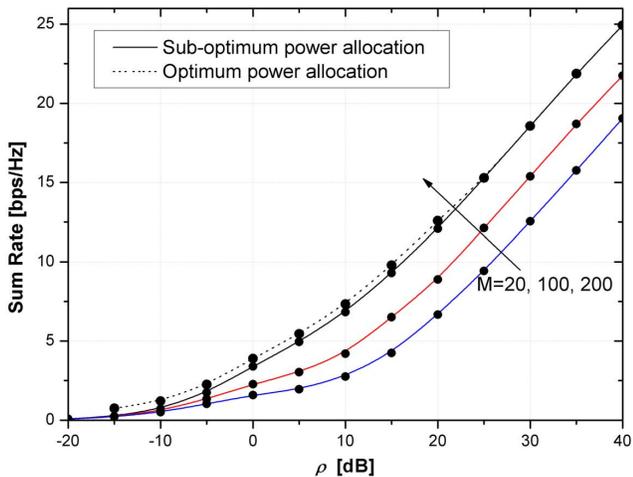}
 \caption{\footnotesize{Sum-rate versus $\rho$ for varying $M$ ($K=10$, $T=500$, $p_{\mathrm{tr}}=2~\mathrm{dB}$).}}
 \label{VaryingM}
 \end{center}
 \end{figure}

Fig.~\ref{VaryingM} presents the sum-rate with RS versus the SNR for increasing $M$. Clearly, the higher the number of transmit antennas $M$, the less severe the SI becomes. In such case, the sum-rate becomes larger. Moreover, we have added a line in the case that $M=200$ corresponding to the optimum solution, in order to compare the results from the proposed sub-optimum power allocation method and the optimum solution which is obtained numerically. As can be seen, the solution provided by the sub-optimum method appears performance, which is very close to the performance obtained by means of exhaustive search. Especially, in the high-SNR regime, the two lines coincide, which shows that the sub-optimum method behaves as optimum in this region.

\subsubsection{Varying the Number of Destination Users $K$} 
\begin{figure}[!h]
 \begin{center}
 \includegraphics[width=0.95\linewidth]{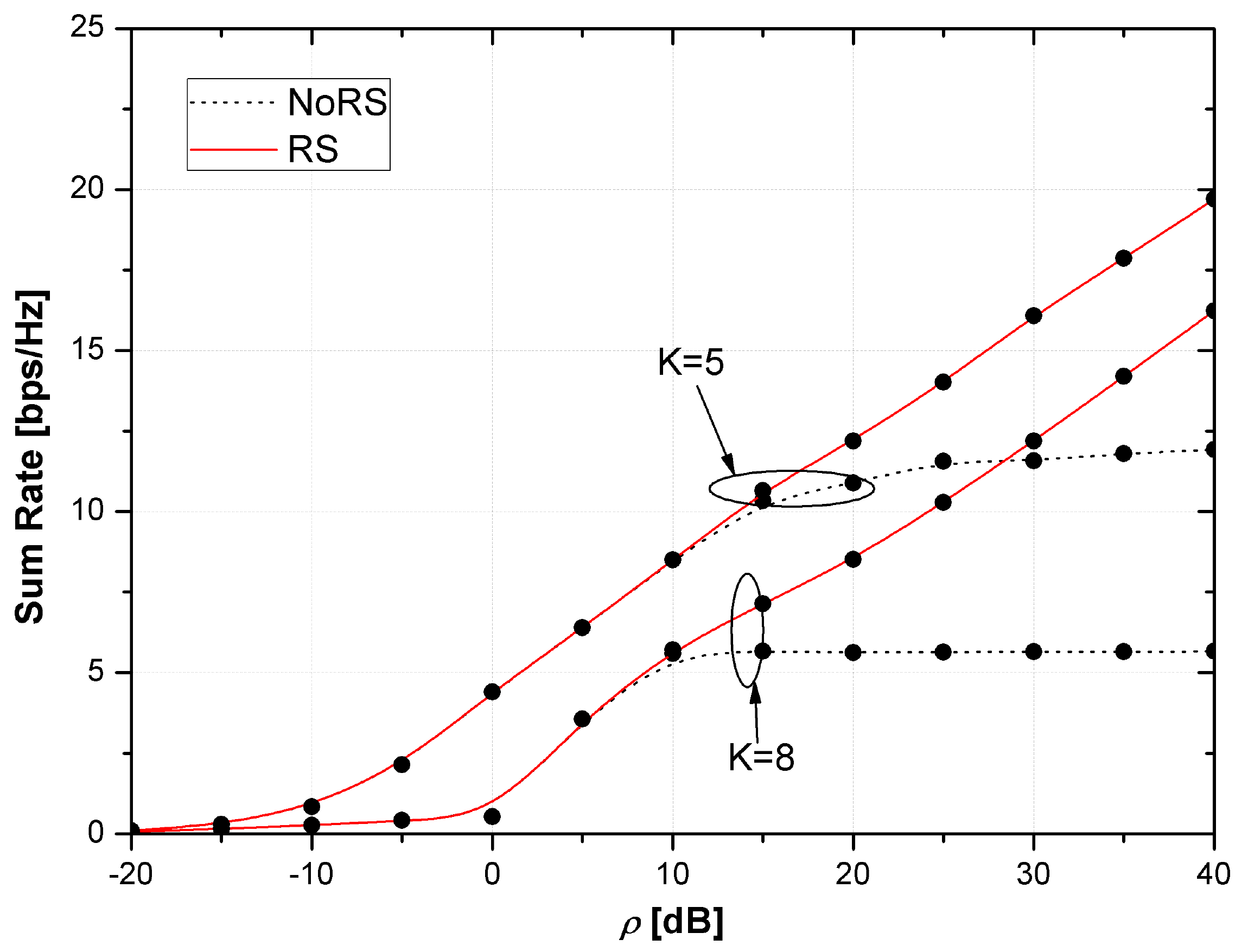}
 \caption{\footnotesize{Sum-rate versus $\rho$ for varying $K$ ($M=100$,  $T=500$, $p_{\mathrm{tr}}=2~\mathrm{dB}$).}}
 \label{VaryingK}
 \end{center}
 \end{figure}

Fig.~\ref{VaryingK} aims at the verification that the achievable rate due to common message degrades with the number of destination users $K$, and also quantifies the sum-rate in the case of FD. Obviously, increasing the number $K$ from $5$ to $8$, the achievable rate degrades because the common message has to be decoded by more destination users. A solution for this, known as hierarchical rate-splitting (HRS), retains the benefits of RS, has been presented in~\cite{Dai2016}. However, the study of HRS  in the case of FD is left for future work.
 \section{Conclusions}\label{Conclusions}
RS achieves to mitigate the degradation emerged in multi-user systems with imperfect CSIT.  Motivated by this observation, we proposed the RS strategy to tackle the saturation occured in  multi-pair MIMO relay systems   with imperfect CSIT.   Interestingly, we considered relay stations employing a large number of antennas (massive  MIMO), in order to address the perfomance issues of the forthcoming 5G networks. Specifically, the objective of this work was to examine the potential robustness of the RS transmission method in multi-pair massive MIMO relay systems obeying mostly to the FD design.

Initially, we presented the RS method and the FD approach. Next, by considering realistic channels with imperfect CSIT, we obtained the estimated channels of the first and the second links. Having applied a conventional multi-user uplink design for the first link, we assumed the implementation of RS in the second link. Actually, we provided the DE analysis of the achievable rate for the first link, and then, for the second one, where the precoders for the common and private messages were designed based on RS. Notably, the validation of the analytical results was demonstrated  by means of simulations that depicted that the asymptotic results can be applicable even for  systems of finite dimensions. Remarkably, RS proved to be robust in both cases of HD and FD relaying. Furthermore, among the interesting outcomes of this paper, it was extracted that by increasing the number of relay antennas or by decreasing the severity of SI, RS appears to be more robust. After a certain value of the SI, this property of RS degrades. Furthermore, we showed that in the case of a dense environment with increasing number of users, the ability of RS to tackle SI and multi-user saturation worsens because the common message has to be decoded by more users. As a future work, we plan to focus on the robustness of RS in different system models with altered CSIT. For example, we plan to implement the RS strategy in millimeter wave systems with imperfect CSIT that consider hybrid precoding.

\begin{appendices}
\section{Proof of  Theorem~\ref{theorem:MMSE}}\label{theorem4}
After appropriate substitution of the MMSE decoder and scaling by $\frac{1}{N}$, the desired signal becomes
\begin{align}
\frac{1}{N}\hat{\bg}_{\mathrm{SR},k}^{\H}\hat{\bSigma}_{\mathrm{SR}}\bg_{\mathrm{SR},k}
&=\frac{1}{N} \hat{\bg}_{\mathrm{SR},k}^{\H}\hat{\bSigma}_{\mathrm{SR}} \hat{\bg}_{\mathrm{SR},k} \\
&=\frac{\frac{1}{N}\tr \hat{\bSigma}_{\mathrm{SR},k} \hat{\bD}_{\mathrm{SR},k}}{1+\frac{1}{N}\tr\hat{\bSigma}_{\mathrm{SR},k} \hat{\bD}_{\mathrm{SR},k}}\\
&=\frac{\frac{1}{N}\tr\hat{\bD}_{\mathrm{SR},k} \bT_{\mathrm{SR},k} }{1+\frac{1}{N}\tr\hat{\bD}_{\mathrm{SR},k}\bT_{\mathrm{SR},k} },
\end{align}
where we have used~\cite[Eq.~2.2]{Bai1}, and ${\hat{\bSigma}}_{\mathrm{SR},k}$ is defined as
${\hat{\bSigma}}_{\mathrm{SR},k}\! =\!\!\big(\hat{\bG}_{\mathrm{SR}}\hat{\bG}_{\mathrm{SR}}^\H\! -\!\hat{\bg}_{\mathrm{SR},k}\hat{\bg}_{\mathrm{SR},k}^{\H}\!+\! \bZ_{\mathrm{SR}}\! +\! N\al_{\mathrm{SR}} \Id_N\!\big)\!^{-1}$.
Applying~\cite[Lem. B.26]{Bai2010a} and~\cite[Thm. 1]{Wagner2012}, we have 
$ \frac{1}{N}\hat{\bg}_{\mathrm{SR},k}^{\H}\hat{\bSigma}_{\mathrm{SR}}\hat{\bg}_{\mathrm{SR},k}\asymp \frac{{\delta}_{\mathrm{SR},k}}{1+{\delta}_{\mathrm{SR},k}} \xrightarrow[ N_{\rt} \rightarrow \infty]{\mbox{a.s.}} 0$,
where ${\delta}_{\mathrm{SR},k}=\frac{1}{N}\tr \hat{\bD}_{\mathrm{SR},k}\bT_{\mathrm{SR},k }$.
The DE of the power of the term concerning the estimation error becomes
\begin{align}
\frac{1}{N^{2}}\big|\bg^\H_{\mathrm{SR},k} \hat{\hat{\bSigma}}_{\mathrm{SR}}\tilde{\bee}_{\mathrm{SR},k}\big|^{2}&= \frac{1}{N^{2}}\bigg|\frac{\hat{\bg}^\H_{\mathrm{SR},k} {\hat{\bSigma}}_{\mathrm{SR},k}\tilde{\bee}_{\mathrm{SR},k}}{1+\hat{\bg}^\H_{\mathrm{SR},k} {\hat{\bSigma}}_{\mathrm{SR},k}\hat{\bg}_{\mathrm{SR},k}}\bigg|^{2}\\
\asymp
&  \frac{1}{N}\frac{{\delta}_{\mathrm{SR},k}^{'}}{\Big(1+{\delta}_{\mathrm{SR},k}\Big)^{2}},
\end{align}
where ${\delta}_{\mathrm{SR},k}^{'}=\frac{1}{N}\tr (\bD_{\mathrm{SR},k} - \hat{\bD}_{\mathrm{SR},k})\bT^{'}_{{\bD}_{\mathrm{SR}},k}$ and $\bK={\bD}_{\mathrm{SR},k}-\hat{\bD}_{\mathrm{SR},k}$.
The DE of the term, including the expression of the received AWGN noise, is obtained as
\begin{align}
\frac{1}{N^{2}}\big|\hat{\bg}^\H_{\mathrm{SR},k}{\hat{\bSigma}}_{\mathrm{SR}}\big|^2
&=\frac{1}{N^{2}}\bigg|\frac{\hat{\bg}^\H_{\mathrm{SR},k}{\hat{\bSigma}}_{\mathrm{SR},k}}{1+\hat{\bg}^\H_{\mathrm{SR},k}{\hat{\bSigma}}_{\mathrm{SR},k}\hat{bg}_{\mathrm{SR},k}}\bigg|^2\\
&=\frac{1}{N^{2}}\frac{\hat{\bg}^\H_{\mathrm{SR},k}{\hat{\bSigma}}^{2}_{\mathrm{SR},k}\hat{\bg}_{\mathrm{SR},k}}{\Big(1+\frac{1}{N}\hat{\bg}^\H_{\mathrm{SR},k}{\hat{\bSigma}}_{\mathrm{SR},k}\hat{\bg}_{\mathrm{SR},k}\Big)^2}\nn\\
& \asymp \frac{1}{N}\frac{{\delta}_{\mathrm{SR},k}^{''}}{\Big(1+{\delta}_{\mathrm{SR},k}\Big)^2},\label{eq:theorem2.I.3}
\end{align}
where ${\delta}_{\mathrm{SR},k}^{''}=\frac{1}{N}\tr \hat{\bD}_{\mathrm{SR},k}\bT_{\mathrm{SR},k}^{''}$ and $\bK=\Id_N$. Note that we have applied \cite[Lem. B.26]{Bai2010a} and~\cite[Thm. 1]{Wagner2012}.
The derivation of the DE of the power of the multi-user interference follows. We have
\begin{align}
&\frac{1}{N^{2}} \big|\hat{\bg}^\H_{\mathrm{SR},k} {\hat{\bSigma}}_\mathrm{SR} \bg_{\mathrm{SR},j} \big|^{2}= \frac{1}{N^{2}}\bigg|\frac{\hat{\bg}^\H_{\mathrm{SR},k} {\hat{\bSigma}}_{\mathrm{SR},k} \bg_{\mathrm{SR},j} }{1+\hat{\bg}^\H_{\mathrm{SR},k} {\hat{\bSigma}}_{\mathrm{SR},k} \hat{\bg}_{\mathrm{SR},k} }\bigg|^{2}\nn\\
&\asymp\frac{1}{N^{2}}\frac{\hat{\bg}^\H_{\mathrm{SR},k} {\hat{\bSigma}}_{\mathrm{SR},k} \bD_{\mathrm{SR},j}{\hat{\bSigma}}_{\mathrm{SR},k} \hat{\bg}_{\mathrm{SR},k} }{\left(1+\hat{\bg}^\H_{\mathrm{SR},k} {\hat{\bSigma}}_{\mathrm{SR},k} \hat{\bg}_{\mathrm{SR},k} \right)^{2}}.\label{eq:theorem2.I.4}
\end{align}
Since ${\hat{\bSigma}}_{\mathrm{SR},k} $ is not independent of $\bg_{\mathrm{SR},j} $, the use of~\cite[Lemma~2]{Hoydis2013} gives
\begin{align}
{\hat{\bSigma}}_{\mathrm{SR},k} \!=\!{\hat{\bSigma}}_{\mathrm{SR},jk} \!-\!\frac{{\hat{\bSigma}}_{\mathrm{SR},jk} \hat{\bg}_{\mathrm{SR},j} \hat{\bg}^\H_{\mathrm{SR},j} {\hat{\bSigma}}_{\mathrm{SR},jk} }{1+\hat{\bg}^\H_{\mathrm{SR},k} {\hat{\bSigma}}_{\mathrm{SR},jk} \hat{\bg}_{\mathrm{SR},k} },\label{eq:theorem2.I.511}
\end{align}which introduces a new matrix ${\hat{\bSigma}}_{\mathrm{SR},jk} $ to~\eqref{eq:theorem2.I.4} defined as
${\hat{\bSigma}}_{\mathrm{SR},jk} =\big(\hat{\bG}_{\mathrm{SR}} \hat{\bG}_{\mathrm{SR}}^\H  -\hat{\bg}_{\mathrm{SR},j} \hat{\bg}^\H_{\mathrm{SR},j} - 
\hat{\bg}_{\mathrm{SR},k} \hat{\bg}^\H_{\mathrm{SR},k}+ \bZ_{\mathrm{SR},k} + N\al_{\mathrm{SR}}  \Id_N\big)^{-1}$.
By substituting~\eqref{eq:theorem2.I.511} into~\eqref{eq:theorem2.I.4} and applying  \cite[Lem. B.26]{Bai2010a} and \cite[Thm. 1]{Wagner2012}, we obtain
 $\frac{1}{N^{2}} \big|\hat{\bg}^\H_{\mathrm{SR},k} \hat{\hat{\bSigma}}_{\mathrm{SR},k} \bg_{\mathrm{SR},j} \big|^{2}
\asymp\frac{{\mu}_{\mathrm{SR},jk}}{N\left(1+\hat{\delta}_{\mathrm{SR},k}\right)^{2}}$,            
where ${\mu}_{\mathrm{SR},jk}$ is given by~\eqref{eq:theorem2.I.mu15}. The derivation of the DE of each term of~\eqref{eq:theorem2.I.mu15} follows.
\begin{figure*}
\begin{align}
{\mu}_{\mathrm{SR},k}&= \hat{\bg}^\H_{\mathrm{SR},k} \hat{\bSigma}_{\mathrm{SR},jk}\hat{\bD}_{\mathrm{SR},j} \hat{\bSigma}_{\mathrm{SR},jk}\hat{\bg}_{\mathrm{SR},k}\!+\!\frac{\left|  \hat{\bg}^\H_{\mathrm{SR},k}\hat{\bSigma}_{\mathrm{SR},jk} \hat{\bg}_{\mathrm{SR},k}\right|^{2}\hat{\bg}^\H_{\mathrm{SR},k} \hat{\bSigma}_{\mathrm{SR},jk}\hat{\bD}_{\mathrm{SR},j}\hat{\bSigma}_{\mathrm{SR},jk}\hat{\bg}_{\mathrm{SR},k}}{\left( 1+\hatvg^\H_{\mathrm{SR},k} \hat{\bSigma}_{\mathrm{SR},jk} \hatvg_{\mathrm{SR},k} \right)^{2}}\nn\\
&-2\mathrm{Re}\left\{  \frac{\hatvg^\H_{\mathrm{SR},k}\hat{\bSigma}_{\mathrm{SR},jk} \bg_{\mathrm{SR},k}\bg_{\mathrm{SR},k}^{\H}\hat{\bSigma}_{\mathrm{SR},jk}\hat{\bD}_{\mathrm{SR},j}\hat{\bSigma}_{\mathrm{SR},jk}\hatvg_{\mathrm{SR},k}}{1+\hatvg^\H_{\mathrm{SR},k} \hat{\bSigma}_{\mathrm{SR},jk} \hatvg_{\mathrm{SR},k}}\right\}.
 \label{eq:theorem2.I.mu15}
\end{align}
\line(1,0){470}
\end{figure*}
In particular, we have
\begin{align}
 \frac{1}{N^{2}}\hat{\bg}^\H_{\mathrm{SR},k} \hat{\bSigma}_{\mathrm{SR},jk}
 &\hat{\bD}_{\mathrm{SR},j} \hat{\bSigma}_{\mathrm{SR},jk} \hat{\bg}_{\mathrm{SR},k} \nn\\ &\asymp  \frac{1}{N^{2}}\tr \hat{\bD}_{\mathrm{SR},k}\hat{\bSigma}_{\mathrm{SR},jk}\hat{\bD}_{\mathrm{SR},j}\hat{\bSigma}_{\mathrm{SR},jk}\nn\\
 &\asymp\frac{1}{N^{2}}\tr \hat{\bD}_{\mathrm{SR},j}{\bT}_{\mathrm{SR},jk}^{'} =\frac{ \delta_{\mathrm{SR},jk}^{'}}{N},
\end{align}
where we have used \cite[Thm. 1]{Wagner2012} and \cite[Thm. 2]{Hoydis2013} as well as we have defined $ \delta_{\mathrm{SR},jk}'=\frac{1}{N}\tr \hat{\bD}_{\mathrm{SR},j}{\bT}_{\mathrm{SR},jk}^{'}$.
Moreover, we have used $\hat{\bg}^\H_{\mathrm{SR},k} \hat{\hat{\bSigma}}_{\mathrm{SR},jk} \hat{\bg}_{\mathrm{SR},k} \asymp\frac{1}{N}\tr\hat{\bD}_{\mathrm{SR},k}\mathrm{\bT}_{\mathrm{SR},jk}=\delta_{\mathrm{SR},jk}$ by means of \cite[Thm. 1]{Wagner2012} and \cite[Thm. 2]{Hoydis2013}.
As shown above, application of Lemma~\cite[Lem. B.26]{Bai2010a} as well as~\cite[Thm. 1]{Wagner2012} and~\cite[Thm. 2]{Hoydis2013} to the first and second term of~\eqref{eq:theorem2.I.mu15} gives
\begin{align}
&\frac{1}{N^{2}}\!\left|  \hat{\bg}^\H_{\mathrm{SR},k}\hat{\bSigma}_{\mathrm{SR},jk} \hat{\bg}_{\mathrm{SR},k}\right|^{2}
\!\asymp \!\frac{1}{N^{2}}\!\tr\hat{\bD}_{\mathrm{SR},k}\hat{\bSigma}_{\mathrm{SR},jk}\hat{\bD}_{\mathrm{SR},k}\hat{\bSigma}_{\mathrm{SR},jk}\nn\\
&\asymp \frac{1}{N^{2}}\tr\hat{\bD}_{\mathrm{SR},k}{\bT}_{\mathrm{SR},jk} ^{''} =\frac{ \delta_{\mathrm{SR},jk}^{''}}{N}.
\end{align}
Regarding the term including the SI, we have
\begin{align}
 &\frac{1}{N^{2}}\bw_{k}^{\H}\bG_{\mathrm{RR}}\EE\left[ \bs\left[n \right]\bs^{\H}\left[n \right]\right] \bG_{\mathrm{RR}}^{\H} \bw_{k}=\frac{1}{N^{2}}\bw_{k}^{\H}\bG_{\mathrm{RR}} \bG_{\mathrm{RR}}^{\H} \bw_{k}\nn\\
 &=\frac{1}{N^{2}}
 \hatvg^\H_{\mathrm{SR},k}\hat{\bSigma}_{\mathrm{SR}}
 \bG_{\mathrm{RR}}\bG_{\mathrm{RR}}^{\H}\hat{\bSigma}_{\mathrm{SR}}\hatvg_{\mathrm{SR},k}\\
 &=\frac{1}{N}\frac{\frac{1}{N}\hatvg^\H_{\mathrm{SR},k}\hat{\bSigma}_{\mathrm{SR},k}\bG_{\mathrm{RR}}\bG_{\mathrm{RR}}^{\H}\hat{\bSigma}_{\mathrm{SR},k}\hatvg_{\mathrm{SR},k}}{\left(1+ \frac{1}{N}\hatvg^\H_{\mathrm{SR},k}\hat{\bSigma}_{\mathrm{SR},k}\hatvg_{\mathrm{SR},k} \right)^{2}}\label{eq3} \\
 &\asymp \frac{1}{N}\frac{\frac{1}{N}\tr \bD_{\mathrm{SR},k}\hat{\bSigma}_{\mathrm{SR},k}\bG_{\mathrm{RR}}\bG_{\mathrm{RR}}^{\H}\hat{\bSigma}_{\mathrm{SR},k}}{\left(1+ \frac{1}{N}\tr \bD_{\mathrm{SR},k}\hat{\bSigma}_{\mathrm{SR},k} \right)^{2}}\label{eq4}\\
 &\asymp \frac{1}{N}\frac{\frac{1}{N}\tr \bD_{\mathrm{SR},k}\bT^{''}_{\mathrm{SR},k}
 \frac{1}{N}\tr \bG_{\mathrm{RR}}\bG_{\mathrm{RR}}^{\H} }{\left(1+ \frac{1}{N}\tr \bD_{\mathrm{SR},k}\bT_{\mathrm{SR},k}\right)^{2}}\label{eq5}\\
  &=  \frac{\delta_{\mathrm{SR},jk}^{''}
 \frac{1}{N}\tr \bT_{\mathrm{RR}}  }{\left(1+  \delta_{\mathrm{SR},k}\right)^{2}},
\end{align} where in \eqref{eq3}, we have used \cite[Eq.~2.2]{Bai1} twice. Next, in \eqref{eq4}, we have applied \cite[Lem. B.26]{Bai2010a}, while in~\eqref{eq5}, \cite[Eq.~2.2]{Bai1} is used. Note that $\frac{1}{N}\tr \bT_{\mathrm{RR}}=\frac{1}{N}\tr \bG_{\mathrm{RR}}\bG_{\mathrm{RR}}^{\H}$ due to Theorem 
  with the covariance of the $i$th column of $\bG_{\mathrm{RR}}$ equal to $\sigma^{2}_{\mathrm{SI}}\Id_{N}$. If we  make the necessary substitutions, we obtain the corresponding deterministic equivalent $\overbar{\gamma}_{\mathrm{SR},k}$ and this concludes the proof.

\section{Proof of  Theorem~\ref{theorem:RZF}}\label{theorem3}
First, we derive the DE of the normalization parameter $\lambda$. We  start with a simple algebraic manipulation to $\lambda$. We obtain
$\lambda=
\frac{K}{\tr  \hat{\bG}^{\H}_{\mathrm{RD}}{ \hat{\bSigma}}_{\mathrm{RD}}^{2}\hat{\bG}_{\mathrm{RD}} } = \frac{K}{\Psi}$.
 Next, we have
\begin{align}
 \Psi&=\sum_{k=1}^{K}\hat{\bg}_{\mathrm{RD},k}^{\H} \hat{\bSigma}^{2}_{\mathrm{RD}}\hat{\bg}_{\mathrm{RD},k}\\
 &\asymp\frac{1}{M}\sum_{k=1}^{K}\frac{\frac{1}{M}\tr \hat{\bD}_{\mathrm{RD},k}\hat{\bSigma}_{\mathrm{RD},k}^{2}}{\left( 1+\frac{1}{M}\tr \hat{\bD}_{\mathrm{RD},k}\hat{\bSigma}_{\mathrm{RD},k} \right)^{2}}\label{lambda1}\\
 &\asymp\frac{1}{M}\sum_{k=1}^{K}\frac{{\delta}_{\mathrm{RD},k}^{'}}{\left( 1+{\delta}_{\mathrm{RD},k} \right)^{2}},\label{lambda2} =\frac{K}{\bar{\lambda}},
\end{align}
where we have defined $
 {\hat{\bSigma}}_{\mathrm{RD},k}\!\triangleq\!\left(\! \hat{\bW}_{\mathrm{RD}} \!-\!\hat{\bg}_{\mathrm{RD},k}\hat{\bg}_{\mathrm{RD},k}^{\H} \! +\! {\bZ_{\mathrm{RD}}} \!+\!  \al_{\mathrm{RD}} M \Id_M\right)^{\!\!-1}$ with $\hat{\bW}_{\mathrm{RD}}=\hat{\bG}_{\mathrm{RD}}\hat{\bG}_{\mathrm{RD}}^\H$.
Also, we have applied~\cite[Thm. 1]{Wagner2012} and~\cite[Thm. 2]{Hoydis2013} for $\bL=\hat{\bD}_{\mathrm{RD},k}$ and $\bK=\Id_M$, and we have denoted ${\delta}_{\mathrm{RD},k}=\frac{1}{M}\tr \hat{\bD}_{\mathrm{RD}k}\bT_{\mathrm{RD},k}$ and $\delta_{\mathrm{RD},k}^{'}=\frac{1}{M}\tr \hat{\bD}_{\mathrm{RD},k}{\bT}_{\mathrm{RD},k}^{'}$. Hence, ${\lambda}\asymp \bar{\lambda} $.
Regarding the rest part of the desired signal power, we substitute the RZF precoder, and after dividing by $\frac{1}{M}$ we have
\begin{align}
\frac{1}{M}{\bg}_{\mathrm{RD},k}^{\H}\hat{\bSigma}_{\mathrm{RD}} \hat{\bg}_{\mathrm{RD},k}
&=\frac{1}{M} \hat{\bg}_{\mathrm{RD},k}^{\H}\hat{\bSigma}_{\mathrm{RD}} \hat{\bg}_{\mathrm{RD},k}\label{desired2} \\
&=\frac{\frac{1}{M}\tr \hat{\bSigma}_{\mathrm{RD},k} \hat{\bD}_{\mathrm{RD},k}}{1+\frac{1}{M}\tr\hat{\bSigma}_{\mathrm{RD},k} \hat{\bD}_{\mathrm{RD},k}}\label{desired3}\\
&=\frac{\frac{1}{M}\tr\hat{\bD}_{\mathrm{RD},k} \bT_{\mathrm{RD},k} }{1+\frac{1}{M}\tr\hat{\bD}_{\mathrm{RD},k}\bT_{\mathrm{RD},k} },\label{desired5}
\end{align}
 where we have applied Lemmas~\cite[Eq.~2.2]{Bai1},~\cite[Lem. B.26]{Bai2010a}, and \cite[p. 207]{Tao2012} in \eqref{desired2} and \eqref{desired3}, respectively. Moreover, we have exploited~\cite[Thm. 1]{Wagner2012} in  \eqref{desired5} for $\bL=\hat{\bD}_{\mathrm{RD},k}$. 
Writing in  a concise form the  last equation, we obtain
$\frac{1}{M}\bg_{\mathrm{RD},k}^{\H}\hat{\bSigma}_{\mathrm{RD}} \hat{\bg}_{\mathrm{RD},k}=\frac{ {\delta}_{\mathrm{RD},k}}{1+{\delta}_{\mathrm{RD},k} }$.
We continue the proof with the derivation of the DE of the term of the interference part of 
$\frac{\rho_{j}}{K}\sum_{j\ne k}^{K}|{\bg}_{\mathrm{RD},k}^{\H}\bff_{\mathrm{RD},j}|^2$.
Making use of \cite[Eq.~2.2]{Bai1}, we obtain by means of~\cite[Lem. B.26]{Bai2010a} and~\cite[Eq.~2.2]{Bai1}
\begin{align}
&\frac{1}{M^{2}}|{\bg}_{\mathrm{RD},k}^{\H}\bff_{\mathrm{RD},j}|^2=\frac{1}{M^{2}}|\hat{\bg}_{\mathrm{RD},k}^{\H}\hat{\bSigma}_{\mathrm{RD}} \hat{\bg}_{\mathrm{RD},j}|^2\label{Int1} \\
&=\frac{1}{M^{2}}\frac{\hat{\bg}_{\mathrm{RD},k}^{\H}\hat{\bSigma}_{\mathrm{RD},j} \hat{\bg}_{\mathrm{RD},j}\hat{\bg}^{\H}_{\mathrm{RD},j}\hat{\bSigma}_{\mathrm{RD},j}\hat{\bg}_{\mathrm{RD},k}}{\left( 1+\hat{\bg}_{\mathrm{RD},j}^{\H}\hat{\bSigma}_{\mathrm{RD},j} \hat{\bg}_{\mathrm{RD},j} \right)^2}\label{Int2}\\
&=\frac{1}{M^{2}}\frac{\hat{\bg}_{\mathrm{RD},k}^{\H}\hat{\bSigma}_{\mathrm{RD},j} \hat{\bD}_{\mathrm{RD},j}\hat{\bSigma}_{\mathrm{RD},j}\hat{\bg}_{\mathrm{RD},k}}{\left( 1+\hat{\bg}_{\mathrm{RD},j}^{\H}\hat{\bSigma}_{\mathrm{RD},j} \hat{\bg}_{\mathrm{RD},j} \right)^2}\label{Int3},
\end{align}
where in~\eqref{Int3}, we have taken into consideration that $\hat{\bg}_{\mathrm{RD},k} $ and $\hatvg_{\mathrm{RD},j} $ are mutually independent. Given that $\hat{\bSigma}_{\mathrm{RD},j}$  is not independent of $\hat{\bg}_{\mathrm{RD},k}$, we employ ~\cite[Lemma~2]{Hoydis2013}, which yields
\begin{align}
\hat{\bSigma}_{\mathrm{RD},j}={\hat{\bSigma}}_{\mathrm{RD},jk}-\frac{{\hat{\bSigma}}_{\mathrm{RD},jk}\hatvg_{\mathrm{RD},k}\hatvg_{\mathrm{RD},k}^{\H}{\hat{\bSigma}}_{\mathrm{RD},jk}}{1+\hatvg^\H_{\mathrm{RD},k}  {\hat{\bSigma}}_{\mathrm{RD},jk}\hatvg_{\mathrm{RD},k}}\label{eq:theorem2.I.51}, 
\end{align}
where the new matrix ${\hat{\bSigma}}_{\mathrm{RD},jk}$ is defined as
${\hat{\bSigma}}_{\mathrm{RD},jk}\!=\!
\big(\!\hat{\bW}_{\mathrm{RD}} \!-\!\hat{\bg}_{\mathrm{RD},k}\hat{\bg}_{\mathrm{RD},k}^{\H}\!-\!\hat{\bg}_{\mathrm{RD},j}\hat{\bg}_{\mathrm{RD},j}^{\H} \nn\\
+\!  {\bZ_{\mathrm{RD}}} \!+\!  \al_{\mathrm{RD}}~\! \Id_M\big)^{-1}$.
After substituting~\eqref{eq:theorem2.I.51} into~\eqref{Int3}, we have
  $\frac{1}{M^{2}}\left|\hat{\bg}^\H_{\mathrm{RD},k}{\hat{\bSigma}}_{\mathrm{RD}} \hatvg_{\mathrm{RD},j} \right|^{2} 
=\frac{{Q}_{\mathrm{RD},jk}}{M\left(1+{\delta_{\mathrm{RD},k}}\right)^{2}}$,           
where ${Q}_{\mathrm{RD},jk}$ is given by~\eqref{eq:theorem2.I.mu1}.
We proceed with the derivation of the DE of each term in~\eqref{eq:theorem2.I.mu1}. Specifically, we have
\begin{align}
 \frac{1}{M^{2}}\hat{\bg}^\H_{\mathrm{RD},k}& \hat{\bSigma}_{\mathrm{RD},jk}\hat{\bD}_{\mathrm{RD},j} \hat{\bSigma}_{\mathrm{RD},jk} \hat{\bg}_{\mathrm{RD},k}\nn\\
 &
 \asymp  \frac{1}{M^{2}}\tr \hat{\bD}_{\mathrm{RD},k}\hat{\bSigma}_{\mathrm{RD},jk}\hat{\bD}_{\mathrm{RD},j}\hat{\bSigma}_{\mathrm{RD},jk}\\
 &\asymp\frac{1}{M^{2}}\tr \hat{\bD}_{\mathrm{RD},j}{\bT}_{\mathrm{RD},jk}^{''} =\frac{ \delta_{\mathrm{RD},jk}^{''}}{M},
\end{align}
\begin{figure*}
\begin{align}
{Q}_{\mathrm{RD},jk}&= \hat{\bg}^\H_{\mathrm{RD},k} \hat{\bSigma}_{\mathrm{RD},jk}\hat{\bD}_{\mathrm{RD},j} \hat{\bSigma}_{jk}\hat{\bg}_{\mathrm{RD},k}\!+\!\frac{\left|  \hat{\bg}^\H_{\mathrm{RD},k}\hat{\bSigma}_{\mathrm{RD},jk} \hat{\bg}_{\mathrm{RD},k}\right|^{2}\hat{\bg}^\H_{\mathrm{RD},k} \hat{\bSigma}_{\mathrm{RD},jk}\hat{\bD}_{\mathrm{RD},j}\hat{\bSigma}_{\mathrm{RD},jk}\hat{\bg}_{\mathrm{RD},k}}{\left( 1+\hatvg^\H_{\mathrm{RD},k} \hat{\bSigma}_{\mathrm{RD},jk} \hatvg_{\mathrm{RD},k} \right)^{2}}\nn\\
&-2\mathrm{Re}\left\{  \frac{\hatvg^\H_{\mathrm{RD},k}\hat{\bSigma}_{\mathrm{RD},jk} \bg_{\mathrm{RD},k}\bg_{\mathrm{RD},k}^{\H}\hat{\bSigma}_{\mathrm{RD},jk}\hat{\bD}_{\mathrm{RD},j}\hat{\bSigma}_{\mathrm{RD},jk}\hatvg_{\mathrm{RD},k}}{1+\hatvg^\H_{\mathrm{RD},k} \hat{\bSigma}_{\mathrm{RD},jk} \hatvg_{\mathrm{RD},k}}\right\}.
 \label{eq:theorem2.I.mu1}
\end{align}
\line(1,0){470}
\end{figure*}where we have applied~\cite[Lem. B.26]{Bai2010a} and~\cite[Thm. 2]{Hoydis2013} for $\bL=\hat{\bD}_{\mathrm{RD},k}$ and $\bK=\hat{\bD}_{\mathrm{RD},j}$.
Similarly, we have
\begin{align}
\frac{1}{M^{2}}&\left|  \hat{\bg}^\H_{\mathrm{RD},k}\hat{\bSigma}_{\mathrm{RD},jk} \hat{\bg}_{\mathrm{RD},k}\right|^{2}\nn\\
&\asymp \frac{1}{M^{2}}\tr\hat{\bD}_{\mathrm{RD},k}\hat{\bSigma}_{\mathrm{RD},jk}\hat{\bD}_{\mathrm{RD},k}\hat{\bSigma}_{\mathrm{RD},jk}\nn\\
&= \frac{1}{M^{2}}\tr\hat{\bD}_{\mathrm{RD},k}\hat{\bSigma}_{\mathrm{RD},jk}\hat{\bD}_{\mathrm{RD},k}\hat{\bSigma}_{\mathrm{RD},jk}\\
&\asymp \frac{1}{M^{2}}\tr\hat{\bD}_{\mathrm{RD},k}{\bT}_{\mathrm{RD},jk}^{'''} =\frac{ \delta_{\mathrm{RD},jk}^{'''}}{M},
\end{align}
where $\bL=\hat{\bD}_{\mathrm{RD},k}$ and $\bK=\hat{\bD}_{\mathrm{RD},k}$, and  $\delta_{\mathrm{RD},jk}^{'''}=\frac{1}{M}\tr \hat{\bD}_{\mathrm{RD},jk}{\bT}^{'''}_{\mathrm{RD},jk}$.
 The next term is written as
\begin{align}
&\frac{1}{M^{2}}\hatvg^\H_{\mathrm{RD},k}\hat{\bSigma}_{\mathrm{RD},jk} \bg_{\mathrm{RD},k} \asymp \frac{1}{M}\tr\hat{\bD}_{\mathrm{RD},k}\hat{\bSigma}_{\mathrm{RD},jk}\label{int4} \\
&\asymp  \frac{1}{M}\tr\hat{\bD}_{\mathrm{RD},k}\bT_{\mathrm{RD},jk} =  {\delta}_{\mathrm{RD},jk}\label{int6}, 
\end{align}
where in~\eqref{int4}, we have applied both Lemmas~\cite[Lem. B.26]{Bai2010a} and~\cite[p. 207]{Tao2012}, while in the next equation, we have applied~\cite[Thm. 1]{Wagner2012}. Hence,~\eqref{eq:theorem2.I.mu1} becomes
\begin{align}
\!\!{Q}_{\mathrm{RD},jk}\!&\asymp\! \frac{ \delta_{\mathrm{RD},j}^{'}}{M}\!\!+\!\frac{\left|{\delta_{\mathrm{RD},k}^{''}}\right|^{2}\delta_{\mathrm{RD},k}^{'}}{M\left( 1\!+\!\delta_{\mathrm{RD},k} \right)^{2}}\!\nn\\
&-\!2\mathrm{Re}\left\{ \! \frac{ {\delta}_{\mathrm{RD},jk}^{''}\delta_{\mathrm{RD},k}^{'} }{M\left( 1\!+\!\delta_{\mathrm{RD},k} \right)}\!\right\}\!.
\end{align}
Making the necessary substitutions, the proof is concluded.
\section{Proof of Proposition~\ref{prop:inequality}}\label{proofinequality}
The private messages of the RS transmission in the second link achieve almost the same sum rate as the conventional BC with full power, when $\mathrm{R}_{{\mathrm{RD},k}}^{\mathrm{p}}=\mathrm{R}_{{\mathrm{RD},k}}^{\mathrm{NoRS}}$ holds. Actually, this happens, when 
\begin{align}
\frac{\bar{\lambda}\frac{\rho t}{K}\left(  {\delta}_{\mathrm{RD},k} \right)^{2}}{\bar{\lambda}\frac{\rho t}{K}\displaystyle \sum_{j\ne k}^{K}\frac{{Q}_{\mathrm{RD},jk}}{M}+\left(1+{\delta_{\mathrm{RD},k}}\right)^{2}}>1.
\end{align}
Then, thinking that the number
of users $K$ is generally much larger than $1$, we set
\begin{align}
\frac{\bar{\lambda}\frac{\rho t}{K}\left(  {\delta}_{\mathrm{RD},k} \right)^{2}}{\bar{\lambda}\frac{\rho t}{K}\displaystyle \sum_{j\ne k}^{K}\frac{{Q}_{\mathrm{RD},jk}}{M}+\left(1+{\delta_{\mathrm{RD},k}}\right)^{2}}=K. \end{align}
Another reason for the dependence of this setting on $K$ follows. Specifically, let us denote 
\begin{align}
 \bar{Y}=\frac{\bar{\lambda}\frac{1}{K}\left(  {\delta}_{\mathrm{RD},k} \right)^{2}}{\bar{\lambda}\frac{ 1}{K}\displaystyle \sum_{j\ne k}^{K}\frac{{Q}_{\mathrm{RD},jk}}{M}+\left(1+{\delta_{\mathrm{RD},k}}\right)^{2}}.
\end{align}
Since the common message should be decoded by all destination users, less power is allocated at the common message as  $K$ increases, and equivalently, the rate of the common message reduces. Thus, $\bar{Y}$ should depend on $K$. 

The power  splitting ratio $t$ is chosen as $t=K/\left(
\rho \bar{Y} \right)$. If the choice of  $t$   is the smallest value between $t=K/\left(
\rho \bar{Y} \right)$ and   $1$, the inequality $\mathrm{R}_{{\mathrm{RD},k}}^{\mathrm{p}}\le\mathrm{R}_{{\mathrm{RD},k}}^{\mathrm{NoRS}}$ becomes equality. Note that at low SNR ($\rho \to 0$), $t$  becomes $1$, which means that transmission of the common message is not beneficial, while increasing the SNR, the common message boosts the sum-rate. Similar to~\cite{Dai2016}, the rate loss in the second link between the private messages of the NoRS and RS  is proved to be upper bounded as
\begin{align}
\sum_{j=1}^{K} \left( \mathrm{R}_{\mathrm{RD},j}^{\mathrm{NoRS}}-\mathrm{R}_{\mathrm{RD},j}^{\mathrm{p}} \right)
&\le \log_2 e.\label{log1} 
\end{align}
Substituted \eqref{log1} into \eqref{DR}, we obtain the desired result.

\end{appendices}                                                                                                                                                                                                                                                                        
\section*{Acknowledgement}
The authors would like to express their gratitude to Dr. Bruno Clerckx for his help and support in making this work possible.

\bibliographystyle{IEEEtran} 

\bibliography{mybib}
\end{document}

%% file: def.tex
\usepackage{calc}

\usepackage{xcolor,multirow,gensymb}
  \usepackage{cite}
\usepackage{graphicx,subfigure,epsfig}
\usepackage{psfrag}
\usepackage{amsmath,amssymb}
\usepackage{color}
\usepackage{array}
\usepackage{nicefrac}
\usepackage{amsmath,amssymb}
\usepackage{xifthen}
\usepackage{mathtools}
\usepackage{enumerate}
\usepackage{microtype}
\usepackage{xspace}
\usepackage{bm}
\usepackage[T1]{fontenc}
\usepackage{fancyhdr}
\usepackage{lastpage}
\usepackage{rotating}
\usepackage{tabulary}

\newcommand{\vect}[1]{{\lowercase{\mbs{#1}}}}

\newcommand{\mbs}[1]{\bm{#1}}
\newcommand{\mat}[1]{{\uppercase{\mbs{#1}}}}
\newcommand{\Id}{\mat{\mathrm{I}}}

\newcommand{\T}{{\scriptscriptstyle\mathsf{T}}}
\renewcommand{\H}{{\scriptscriptstyle\mathsf{H}}}

\renewcommand{\Re}[1][]{\ifthenelse{\isempty{#1}}{\operatorname{Re}}{\operatorname{Re}\left(#1\right)}}
\renewcommand{\Im}[1][]{\ifthenelse{\isempty{#1}}{\operatorname{Im}}{\operatorname{Im}\left(#1\right)}}
\usepackage{float}

\newfloat{longequation}{b}{ext}

\newcommand{\ev}{\vect{e}}

\newcommand{\deltav}{\hbox{\boldmath$\delta$}}

\def\bD{{\mathbf{D}}}
\def\bE{{\mathbf{E}}}
\def\bF{{\mathbf{F}}}
\def\bG{{\mathbf{G}}}
\def\bH{{\mathbf{H}}}

\def\bK{{\mathbf{K}}}
\def\bL{{\mathbf{L}}}

\def\bN{{\mathbf{N}}}

\def\bS{{\mathbf{S}}}
\def\bT{{\mathbf{T}}}

\def\bW{{\mathbf{W}}}

\def\bY{{\mathbf{Y}}}
\def\bZ{{\mathbf{Z}}}
\def\bSigma{{\mathbf{\Sigma}}}

\newcommand{\cC}{{\cal C}}

\newcommand{\cN}{{\cal N}}

\def\rt{{\mathrm{t}}}

\def\bb{{\mathbf{b}}}

\def\bee{{\mathbf{e}}}
\def\bff{{\mathbf{f}}}
\def\bg{{\mathbf{g}}}
\def\bh{{\mathbf{h}}}

\def\br{{\mathbf{r}}}
\def\bs{{\mathbf{s}}}

\def\bu{{\mathbf{u}}}

\def\bw{{\mathbf{w}}}

\def\by{{\mathbf{y}}}
\def\bz{{\mathbf{z}}}
\def\b0{{\mathbf{0}}}

\def\bbC{{\mathbb{C}}}

\newcommand{\EE}{\mathbb{E}}

\newcommand{\CN}[1][]{\ifthenelse{\isempty{#1}}{\mathcal{N}_{\mathbb{C}}}{\mathcal{N}_{\mathbb{C}}\left(#1\right)}}

\renewcommand{\P}[1][]{\ifthenelse{\isempty{#1}}{\mathbb{P}}{\mathbb{P}\left(#1\right)}}
\newcommand{\E}[1][]{\ifthenelse{\isempty{#1}}{\mathbb{E}}{\mathbb{E}\left(#1\right)}}
\renewcommand{\det}[1][]{\ifthenelse{\isempty{#1}}{\text{det}}{\text{det}\left(#1\right)}}
\newcommand{\trace}[1][]{\ifthenelse{\isempty{#1}}{\text{tr}}{\text{tr}\left(#1\right)}}
\newcommand{\rank}[1][]{\ifthenelse{\isempty{#1}}{\text{rank}}{\text{rank}\left(#1\right)}}
\newcommand{\diag}[1][]{\ifthenelse{\isempty{#1}}{\text{diag}}{\text{diag}\left(#1\right)}}
\def\nn{\nonumber}

\IEEEoverridecommandlockouts

\newtheorem{proposition}{Proposition}
\newtheorem{remark}{Remark}

\newcommand{\hatvg}{\hat{\bg}} 
 
\def\bPhi{{\boldsymbol{\Phi}}}

\newcommand{\tr}{\mathop{\mathrm{tr}}\nolimits}
\newtheorem{Theorem}{Theorem}

\newcommand{\al}{\alpha}

\newcounter{enumi_saved}
\usepackage{answers}
\Newassociation{solution}{Solution}{solutionfile}

\AtBeginDocument{\Opensolutionfile{solutionfile}[\jobname]}
\AtEndDocument{\Closesolutionfile{solutionfile}\clearpage
}
